\documentclass[12pt]{article}
\usepackage{amsfonts, amssymb, amsmath, epsfig, graphicx}
\newcommand{\A}{\ensuremath{\mathcal{A}}}
\newcommand{\PP}{\ensuremath{\mathcal{P}}}

\begin{document}
\title{Punctured polygons and polyominoes on the square lattice}
\author{Anthony J Guttmann\thanks{e-mail: tonyg@ms.unimelb.edu.au}, 
Iwan Jensen\thanks{e-mail: i.jensen@ms.unimelb.edu.au},
Ling Heng Wong\thanks{e-mail: henryw@ms.unimelb.edu.au}, \\
Department of Mathematics \& Statistics, 
The University of Melbourne,\\
Parkville, Vic. 3010, Australia \\ \\
and Ian G Enting\thanks{e-mail: ige@dar.csiro.au} \\ 
CSIRO, Division of Atmospheric Research, \\
Aspendale, Vic. 3195, Australia}
\date{\today}
\maketitle
\bibliographystyle{plain}
\begin{abstract}
We use the finite lattice method to count the number of punctured
staircase and self-avoiding polygons with up to three holes on the
square lattice. New or radically extended series have been
derived for both the perimeter and area generating functions.
We show that the
critical point is unchanged by a finite number of punctures, and
that the critical exponent increases by a fixed amount for each
puncture. The increase is 1.5 per puncture when enumerating by 
perimeter and 1.0 when enumerating by area. A refined estimate of 
the connective constant for polygons by area is given.
A similar set of results is obtained for finitely punctured polyominoes.
The exponent increase is proved to be 1.0 per puncture for polyominoes.
\end{abstract}

\section{Introduction}

\begin{figure}
\begin{center}
\includegraphics{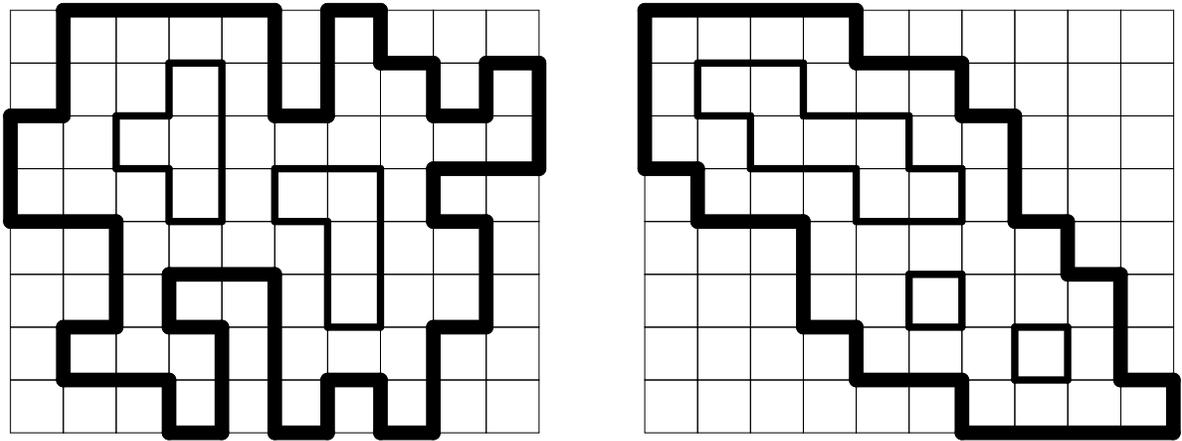}
\end{center}
\caption{\label{fig:puncpol}
An example of a punctured self-avoiding polygon (left panel) and 
a punctured staircase polygon (right panel). The thick lines show the
perimeter of the enclosing polygon while the medium lines show the
perimeter of the holes.}
\end{figure}

A self-avoiding polygon (SAP) can be defined as a walk on a lattice
which returns to the origin and has no other self-intersections.
Alternatively we can define a SAP as a connected sub-graph 
(of a lattice) whose vertices are of degree 2.
The history and significance of this problem is nicely discussed
in \cite{Hug}.  Generally SAP are considered distinct up to a translation, 
so if there are $p_n$ SAP  of length $n$ there are $2np_n$ returns to 
the origin (the factor of two arising
from the two possible directions in which the loop can be traveled).
Staircase polygons are a special case of SAP.
A staircase polygon can be viewed as
the intersection of two directed walks starting at the origin,
taking steps either to the right or down and terminating once
the walks meet. We define a punctured SAP as a 
SAP with one or more holes, with the perimeter of each hole itself being
a SAP. In other words a punctured SAP is a SAP enclosing one or more
SAP. To avoid any possible confusion in our definition of punctured
polygons, note that the
punctures are disjoint --- there are no degree four vertices
in the punctured polygons we are considering.
A similar definition can be given for punctured staircase
polygons. We have shown an example of each case in  
figure~\ref{fig:puncpol}. The two principal questions one can
ask are ``how many polygons, distinct up to a translation, are there
of perimeter $2n$ (including the perimeter of the holes) with $k$ 
punctures?" and ``how many polygons, distinct up to a translation, are 
there of area $n$ with $k$ punctures?"

\begin{figure}
\begin{center}
\includegraphics[height=4cm]{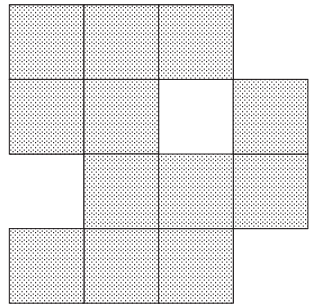}
\caption{\label{fig:polnotsap}
An example of a punctured polyomino that is not a punctured polygon}
\end{center}
\end{figure}

A polyomino is defined as the union of connected cells,
where a cell is a unit square with 4 edges (and its interior).  
Two cells are said to be {\em joined} if they share a common
edge, and are said to be {\em connected} if there exists a sequence of
cells joining the two cells such that successive pairs of cells
are joined.  A $k$-punctured polyomino is a polyomino with $k$ holes.
Unlike the situation for punctured polygons, degree 4 vertices
are permitted, but two holes meeting only at a vertex do count as
two holes, rather than one, as they are not joined.
Punctured polygons are a proper subset of punctured polyominoes. The 
difference between them is that punctured polygons do not include those 
polyominoes where there are 2 cells touching each other only at a vertex.  
An example is shown in figure~\ref{fig:polnotsap}.

Reverting to $k$-punctured polygons, for
$k=0$ we have the simpler --- but still unsolved --- problem
of the number of polygons of perimeter $2n$ or area $n.$ Both these 
problems have been extensively studied for several decades. The most
recent result for polygon perimeters appears to be \cite{JG99} where
polygons of perimeter up to 90 steps are given. In that paper our
analysis of the polygon perimeter generating function led us to
conclude that
$$
P^{(0)}(x)  =  \sum_{n} p^{(0)}_{2n} x^{n} \sim A^{(0)}(x) + 
B^{(0)}(x)(1-\mu^2 x)^{2-\alpha},
$$
where $p^{(0)}_{2n}$ is the number of SAP of perimeter $2n$, distinct 
up to translations. The analysis in \cite{JG99} yielded a very accurate
estimate for the connective constant $\mu = 2.63815853034(10)$ and
confirmed the theoretical prediction $\alpha = 1/2$ \cite{Nienhuis}.
Furthermore, we obtained estimates for the critical amplitudes 
$A^{(0)}(x_c) \approx  0.036$ and $B^{(0)}(x_c) \approx 0.234913,$ 
where $x_c=1/\mu^2$. We also concluded that there was no
evidence for a non-analytic correction-to-scaling exponent, so that
we expect the asymptotic form of the coefficients to behave as:
$$
p^{(0)}_{2n} \sim \mu^{2n} n^{-\frac{5}{2}}[a_1 + a_2/n + a_3/n^2 
+ a_4/n^3 + \cdots].
$$
The connective constant $\mu$ is of course the same as that for 
self-avoiding walks on the same lattice \cite{Hug}.

For polygon areas the most recent published 
work appears to be \cite{EG90} in
which the first 20 terms of the area generating function were given
and analysed. In that work it was found that
$$
A^{(0)}(y)  =  \sum_{n} a^{(0)}_{n} y^{n} \sim C^{(0)}(y) + 
D^{(0)}(y)\log(1-\kappa y),
$$
where $a^{(0)}_{n}$ is the number of SAP of area $n$,  
$\kappa \approx 3.97087,$ and the amplitudes $C^{(0)}$ and $D^{(0)}$ 
were not estimated. Recently this series has been extended to 26 
terms \cite{Web}, but in the present work we have devised a new and 
exponentially faster algorithm, as a result of which we have extended 
the series to 42 terms, and we present an analysis of this longer series. 
The connective constant $\kappa$ is found to be slightly smaller than 
that for the related problem of polyominoes \cite{Gol}. 

In the following, we refer to the boundary of a polygon and its
interior as a {\em disc} and so we will be discussing
punctured discs. An unpunctured disc is a SAP.

For {\em punctured} discs, the basic problem is, analogously,
the calculation of the generating functions

\begin{equation}
P^{(k)}(x)  =  \sum_{n} p_{2n}^{(k)} x^{n} \sim B^{(k)}(x) + 
C^{(k)}(x)(1-(\mu^{(k)})^2 x)^{2-\alpha_k},
\end{equation}
and
\begin{equation}
A^{(k)}(y)  =  \sum_{n} a_{n}^{(k)} y^{n} \sim D^{(k)}(y) + 
E^{(k)}(y)(1- \kappa^{(k)} y)^{-\beta_k},
\end{equation}
\noindent
where the superscript $k$ refers to the number of holes,  or punctures.
From the generating function, one wishes to deduce the asymptotic behaviour, 
believed to be as shown on the r.h.s. of the above equations. The major
problem to be investigated is how the behaviour of $P^{(k)}(x)$ and 
$A^{(k)}(y)$ changes as  $k$ is increased. Previous work \cite{Ren,WRen} 
has been confined to the study of punctured SAP by area. There it was
proved that $\kappa^{(k)} = \kappa^{(0)}=\kappa,$ and that if the exponent 
exists, $\beta_k = \beta_0 + k.$ These results apply more generally to 
punctured surfaces, but in this work we are confining ourselves to two 
dimensions. As far as we are aware, there has been no previous work on the 
problem of the perimeter generating function of punctured discs.

The problem is interesting for several reasons. The effect of a
change in geometry is a much studied topic in lattice statistics,
and our study of the change in perimeter exponent with punctures
seems to be entirely new. It has only been possible by the algorithms
we have designed and implemented, which are exponentially faster than 
pre-existing algorithms. A number of related models have been studied 
previously, such as $c$-animals \cite{Mad88,SotW88} and the behaviour
of prime knots in polygons \cite{Orl96}. $c$-animals are lattice animals
with exactly $c$ cycles. In \cite{SotW88} it was proved that if the
number of such animals $a_n(c) \sim \lambda_c^n n^{\theta_c}$ as
$n \rightarrow \infty$, $c$ fixed, then $\theta_c = \theta_0 + c$
provided that $\theta_0$ exists. It had been previously proved \cite{W83}
that $\lambda_c = \lambda_0.$ The change in connective constant of
$c$-animals as the number of cycles per vertex changes from zero
to non-zero is discussed --- among other results --- in \cite{Mad88}.
Similarly, in a numerical study of knotted polymers \cite{Orl96}, it was 
{\em conjectured} that the exponent $\alpha$ depends on the number of 
prime knots $n_p$ that arise in the knot decomposition of a given SAP via 
the relation $\alpha(n_p) = \alpha(0) + n_p.$

Furthermore, there is considerable pedagogical connection between 
some of these previous studies and our work here, and also between
our work and the study of branched polymers. To sketch this connection,
we first remark that the number of polyominoes, also called lattice 
animals, is just the number of (strongly embedded) site animals. This
connection is readily seen by placing a site at the centre of every cell
of the polyomino, and joining those sites corresponding to joined cells
by a bond. In this way, every lattice animal is mapped to a distinct site 
animal and {\em vice versa.} All other models we are considering map 
similarly to a different subset of strongly embedded site animals. For 
example, SAP map to site animals whose only cycles are 4-cycles (which may 
be isolated or joined). Punctured polygons map to site animals with larger 
cycles, as well as 4-cycles, whereas punctured polyominoes correspond to
site animals with more complex restrictions. Thus this study also
complements the earlier studies of $c$-animals. In those studies,
the variation of exponent with the number of cycles is considered,
whereas in this study we are varying the {\em types} of allowed cycles.

In order to study these and related systems, when an exact solution
can't be found one has to resort to numerical methods. For many problems
the method of series expansions is by far the most powerful method
of approximation. For other problems Monte Carlo methods are superior.
For the analysis of $P^{(k)}(x)$ and $A^{(k)}(y)$, series analysis is 
undoubtedly the most appropriate choice. This method consists of 
calculating the first few coefficients in the expansion of the generating 
function. Given such a series, using the numerical technique known as 
differential approximants \cite{Guttmann89}, highly accurate estimates can 
frequently be obtained for  the critical point and exponents, as well as 
the location and critical exponents of possible non-physical singularities.
Other numerical methods are discussed in \cite{JG99},
and those used in the present study are described more fully below.

In the next section we will describe the finite lattice
method for enumerating punctured polygons. In Section 3 we prove the 
invariance of $\mu^{(k)}$ as $k$ changes, and give an heuristic
argument for the exponent shift with $k.$ The results of
the analysis of the series are presented in Section~\ref{sec:analysis}.
Results analogous to those known for punctured polygons by
area are proved for punctured polyominoes in the Appendix.

\section{Enumeration of punctured polygons \label{sec:flm}}

\subsection{Enumeration of punctured self-avoiding polygons 
\label{sec:sapenum}}

\begin{figure}
\begin{center}
\includegraphics{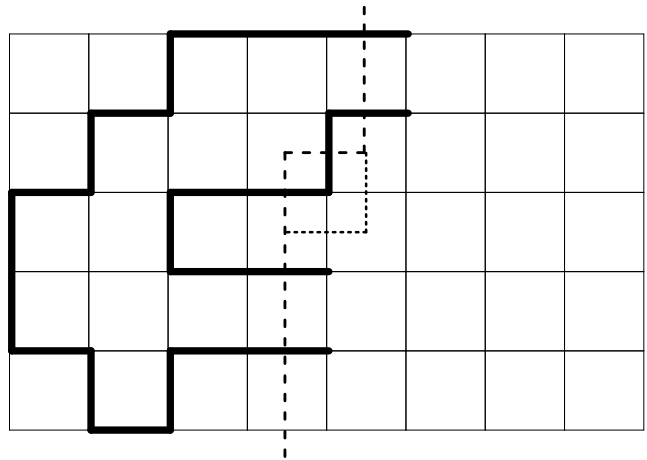}
\end{center}
\caption{\label{fig:transfer}
A snapshot of the intersection (dashed line) during the transfer matrix 
calculation on the square lattice. Polygons are enumerated by successive
moves of the kink in the intersection, as exemplified by the position given 
by the dotted line, so that one vertex at a time is added to the rectangle. 
To the left of the intersection we have drawn an example of a 
partially completed polygon.}
\end{figure}

The method used to enumerate punctured self-avoiding polygons on the 
square lattice is a generalisation of the method devised by 
Enting \cite{Ent} for the enumeration of ordinary SAP. In the following 
we first describe the original method in some detail and show how simple 
it is to generalize the method to the enumeration of punctured polygons.
The first terms in the series for the generating 
function can be calculated using transfer matrix techniques to count 
the number of polygons in rectangles $W+1$ edges wide and $L+1$ edges long. 
The transfer matrix technique  involves drawing a line through the
rectangle intersecting a set of $W+2$ edges. For each configuration of 
occupied or empty edges along the intersection we maintain a (perimeter) 
generating function for loops to the left of the line cutting the 
intersection in that particular pattern. Polygons in a given rectangle 
are enumerated by moving the intersection so as to add one vertex at a time, 
as shown in figure~\ref{fig:transfer}. Since the loops are non-intersecting,
each configuration can be represented by an ordered set of edge states 
$\{n_i\}$, where

\begin{equation*}
n_i  = \left\{ \begin{array}{rl}
 0 &\;\;\; \mbox{empty edge},  \\ 
1 &\;\;\; \mbox{lower part of loop closed to the left}, \\
2 &\;\;\; \mbox{upper part of loop closed to the left}. \\
\end{array} \right.
\end{equation*}
\noindent
Configurations are read from the bottom to the top. So the configuration
along the intersection of the polygon in figure~\ref{fig:transfer} is 
$\{0112122\}$. In passing it is worth noting that there are some major 
restrictions on the possible configurations. Firstly, since all 
loop-ends are connected to the left of the intersection, every lower loop 
end must have a corresponding upper end, and it is therefore clear that the 
total number of `1's  is equal to the total number of `2's. Secondly, 
as we look through the configuration starting from the bottom the 
number of `1's is never smaller than the number of `2's.

\begin{figure}
\begin{center}
\includegraphics{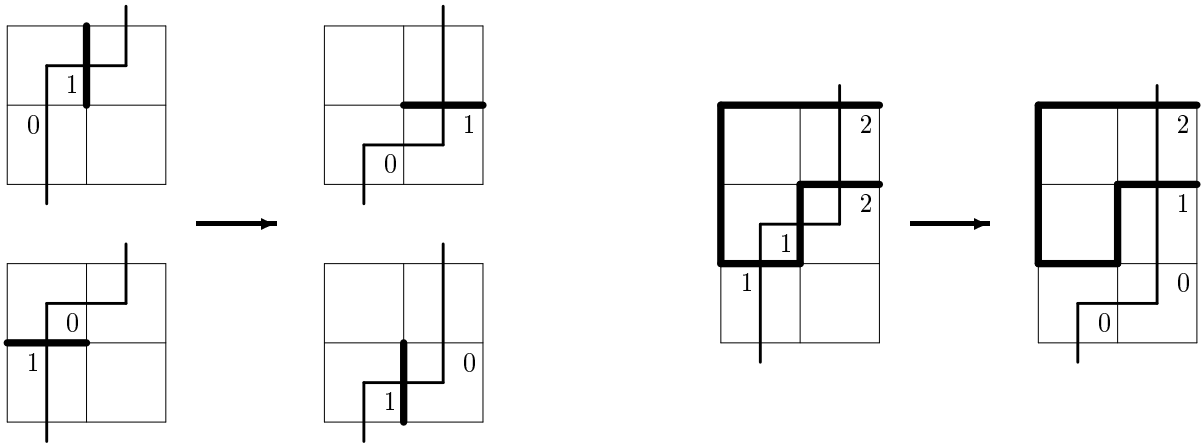}
\end{center}
\caption{\label{fig:update}
Some of the local configurations which occur as the kink in the 
intersection is moved one step. }
\end{figure}

\begin{figure}
\begin{center}
\includegraphics{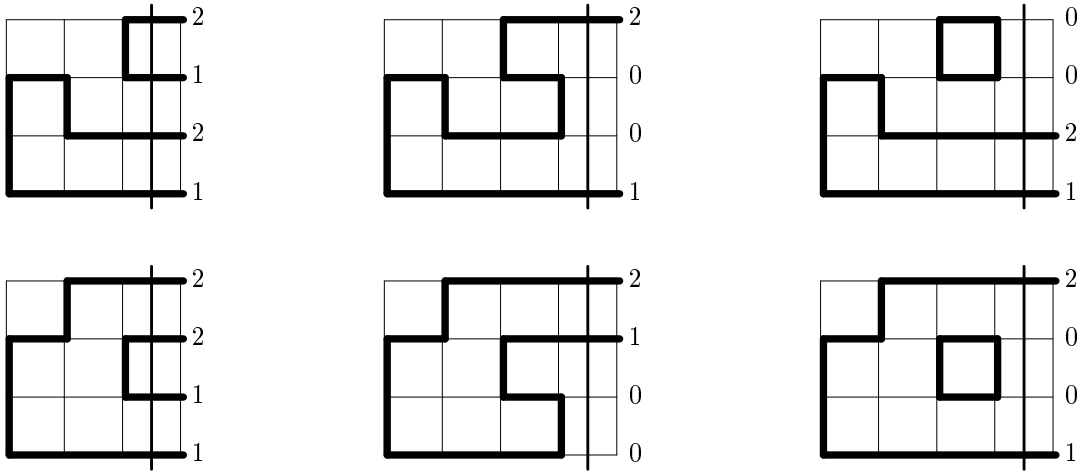}
\end{center}
\caption{\label{fig:sapclose}
Illustration of how a pair of loops can be placed (left), connected
to produce a valid SAP (middle), and connections leading to
forbidden graphs (right).  }
\end{figure}

In Table \ref{tab:update} we have listed the possible local `input' 
states and the `output' states which arise as the kink in the intersection
is propagated by one step. Some of these update rules are illustrated 
further in Fig~\ref{fig:update}. The first panel represents the input 
states `10' and `01' and the possible output states are also `01' and 
`10'. The second panel represents the input state `11' as part of the 
configuration $\{01122\}$. In this case we  connect the two loop ends, 
but in doing so we see that the upper part of the second loop before the 
move becomes the lower part of the one remaining loop after the move. 
That is the configuration $\{01122\}$ becomes $\{00012\}$. This relabelling 
of the other loop-end when connecting two `1's (or two `2's) is denoted 
by over-lining in Table~\ref{tab:update}. In general there could be 
more loops nested in between the two `1's and the corresponding `2' at 
the other end of the loop. Say for instance we had the configuration  
$\{11121222\}$ and connected the first two `1's then the new configuration 
of unconnected loop ends would be  $\{00121212\}$ (drawing a little
figure makes this quite clear). The general rule for the relabelling is
as follows: When connecting two `1's (`2's) we work upward (downward) 
in the configuration, counting the number of `1's and `2's we pass until 
the number of `2's (`1's) exceeds the number of  `1's (`2's). This 
`2' (`1') is the other end of the inner loop and it should now be changed
to a `1' (`2'), thus becoming the lower (upper) end of the outer loop
(again drawing a few pictures should make this clearer).
The weights corresponding to these configuration transformations are simply 
calculated by counting the  number of steps which have been added to the 
polygon. Note that the input state `12' is special because connecting the 
two ends results in a closed loop, so this is only allowed if there are 
no other loops cut by the intersection and the result is a valid polygon, 
which is then accumulated in the total count for that particular length. 
Failure to observe this restriction would result in graphs with 
disconnected components, either one polygon over another or a polygon 
within another (this latter case is of course of interest when we wish
to enumerate punctured polygons). This is illustrated in
figure~\ref{fig:sapclose} where we show the possible ways a pair of loops
can be placed relative to one another (first panels), how the loops
can be connected to produce a valid configuration (middle panels) and
the ways of connecting loops that lead to invalid graphs containing
disconnected components (last panels). In this figure the invalid SAP
in the last panel on the bottom could result in a valid punctured SAP.
Note also that from the input state `00' 
we can produce the output state `12' only if there are other loops
crossing the intersection (otherwise we would produce disconnected
polygons sitting side by side). 
We refer the interested reader to \cite{Ent,EG89} for 
further details regarding the encoding and relabelling of configurations.

Due to the obvious symmetry of the lattice one need only consider 
rectangles with $L \geq W$. In the original approach \cite{Ent} 
valid polygons were required to span the enclosing rectangle in the 
lengthwise direction. So it is clear that polygons with projection on the 
$y$-axis $< W$, that is polygons which are narrower than the width of the 
rectangle, are counted many times. It is however easy to obtain the polygons 
of width exactly $W$ and length exactly $L$ from this enumeration \cite{Ent}. 
Any polygon spanning such a rectangle has a perimeter of length at least 
$2(W+L)$.  By adding the contributions from all rectangles of width 
$W \leq W_{\rm max}$ (where the choice of $W_{\rm max}$ depends on available 
computational resources) and length $W \leq L \leq 2W_{\rm max}-W+1$, 
with contributions from rectangles with $L>W$ counted twice, the number 
of polygons per vertex of an infinite lattice is obtained correctly up 
to perimeter $4W_{\rm max}+2$. 

With the original algorithm the number of configurations required 
as $W_{\rm max}$ increased grew asymptotically as $3^{W_{\rm max}}$
\cite{EG88}. In a recent improvement of the algorithm \cite{JG98,JG99}
valid polygons were required to span the rectangle in 
{\em both} directions. In other words we directly enumerate polygons of
width exactly $W$ and length $L$. For each configuration of
partially completed polygons we keep track of the current minimum 
number of steps $N_{\rm cur}$ that have been inserted to the left of 
the intersection and we calculate the minimum number of additional 
steps $N_{\rm add}$ required to produce a valid polygon that
spans a rectangle of size at least $W \times W$. If the sum 
$N_{\rm cur}+N_{\rm add} > 4W_{\rm max}+2$ the partial generating function 
for that configuration was discarded because it would make no 
contribution to the polygon count up to the perimeter lengths we were
trying to obtain. Numerical evidence indicated that the computational 
complexity was reduced significantly. While the number of configurations 
still grew exponentially as $\lambda^{W_{\rm max}}$ the value of $\lambda$
was reduced from $\lambda = 3$ to $\lambda \simeq 2$ with the 
improved algorithm. Furthermore, for any 
$W$ we know that contributions will start at $4W$ since the smallest 
polygons have to span a $W\times W$ rectangle, so for each configuration 
we need only retain $4(W_{\rm max}-W)+2$ terms of the generating 
functions while in the original algorithm contributions started at 
$2W+2$ because the polygons were required to span only the length-wise 
direction.

The generalization to enumeration of punctured polygons is obtained
by noting that a closed loop is formed whenever we connect a
1-edge to a 2-edge immediately above. If these two edges were 
the only loop ends in the intersection we would have formed a valid
polygon. In other cases we need to ensure that the resulting polygon
is a valid punctured polygon. That is, we must ensure that the
separate SAP just formed will be completely enclosed within a
larger polygon. So we wish to avoid forming separable
polygons, as shown in the upper right panel of figure~\ref{fig:sapclose},
and `holes within holes'. As it turns out the rule for a valid `12' 
closure is simply that we can connect the two loop ends provided there 
is an {\em odd} number of loop ends below the loop being closed. 
To see this consider that as we go through a configuration we note that 
each time we pass a loop end we go from the outside of the polygon to the 
inside and visa versa. So in a configuration all lattice cells
between the first and second loop ends will lie inside the finished
polygon, lattice cells between the second and third loop end will lie
outside the polygon, and so on. Thus we see that by closing a loop
which has an odd number of loop ends below it we are closing off
a part of the lattice which will lie inside the finished polygon.
In particular we see that we avoid the situation shown in the upper right 
panel of figure~\ref{fig:sapclose} with graphs containing
disconnected pieces one over another. Likewise we avoid creating graphs
with disconnected pieces sitting side by side. 

\begin{figure}
\begin{center}
\includegraphics{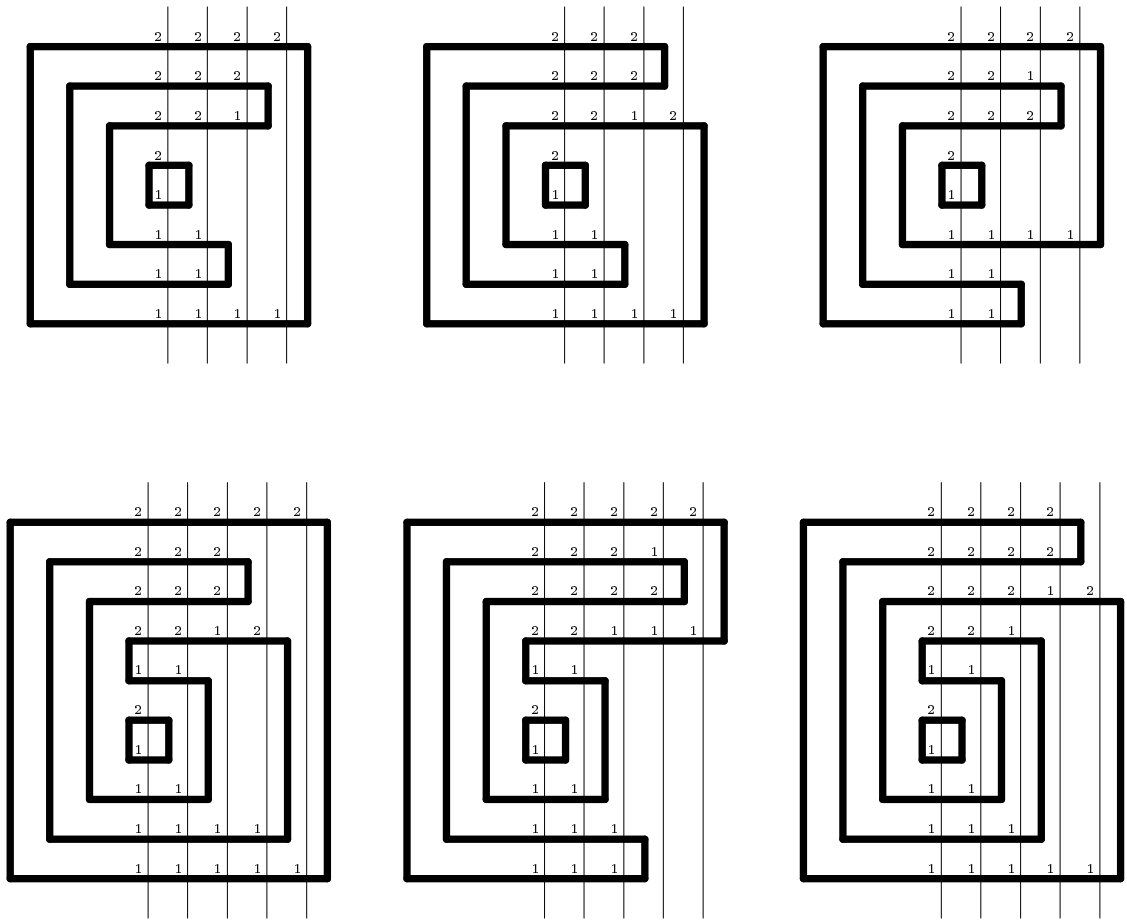}
\end{center}
\caption{\label{fig:puncclose}
Illustration of how connecting loop ends lead to various cases of punctured 
self-avoiding polygons. In the upper (lower) panels we show how the 
configuration \{11112222\} (\{1111212222\}) can be completed given that the
first pair of `12' edges  are connected to form a puncture (in the lower 
panel we further connect the two 1-edges on either side of the puncture). 
In each panel we also show the intersection line with the numbers giving
the labelling of the loop ends.}
\end{figure}

We also avoid forming holes within holes because closing a loop around the 
hole would be prohibited since there would  be an even number of loop ends 
below the `1'-edge (except of course when forming a completed punctured 
polygon). Let us look at the possible edge-configurations around a puncture 
in some detail. First look at the  configuration \{$\ldots 1122 \ldots$\} 
(where the $\ldots$ are any edge configurations with an even number of 
edges making the total configuration of edges a valid intersection). 
The outer `12'-edges can't be connected (the number of edges below the `1' 
is even) so we have to connect two other edges on either side of the hole 
diminishing the number of edges by two and possibly changing the edge labels 
on either side of the hole (in which case we end up with one of the
subsequent cases). In the top panel of figure~\ref{fig:puncclose} we show 
how the simplest interesting case \{11112222\} leads to only valid punctured 
polygons. Secondly, in the case \{$\ldots 1121 \ldots$\}, we can connect
the two `1'-edges forming a partial loop enclosing the hole. In doing so
the matching `2'-edge of the second `1'-edge is changed to a `1'-edge because 
it now is the new lower edge of the larger loop formed by connecting the two 
original `1'-edges. The important thing to notice is that the number of 
edges below this new  `1'-edge is even so we cannot connect it to a `2'-edge 
immediately above and we therefore do not form a hole within a hole. In the 
middle panel of figure~\ref{fig:puncclose} we show the simplest interesting 
case \{1111212222\} and demonstrate that it leads to valid punctured 
polygons. Thirdly, the configuration $\{ \ldots 2121 \ldots$\} is not very
interesting since connecting the `21'-edges in front of the hole
does not result in a loop partially enclosing the hole.
Finally, the configuration \{$\ldots 2122 \ldots$\} is obviously merely
the mirror image of the second case. Note that this covers all cases
since more complicated configurations merely would correspond to more
convoluted loop structures.

In this work we use the generalisation of the algorithm of \cite{JG99}
to count the number of punctured SAPs with up to 3 holes. Obviously, 
the smallest hole we can make in a SAP has perimeter 4 so the number of 
punctured SAPs is obtained correctly up to perimeter $4W_{\rm max}+2+4k$.

The algorithm used for the enumeration of punctured SAPs by area 
is a simple variation of the algorithm described above. The encoding
of configurations along the intersection and the transformation of
these configurations as the intersection is moved remain the same.
The only change is that the weights are different. In order to count
the enclosed area we proceed as follows: A unit of area may be added
as the kink is moved to a new lattice cell (in figure~\ref{fig:transfer}
this is the cell in which the dotted lines meet). Whether or not
a unit of area is added is determined by whether or not
this lattice cell is inside or outside the polygon. But we already
know from the arguments given above that a lattice cell is inside the 
polygon if the number of loop ends below the cell is odd (note that
any loop end along the vertical edge cutting the horizontal part
of the kink is included in the count). So in the case shown in 
figure~\ref{fig:transfer} the lattice cell to which the kink is moved 
lies outside the polygon (there are two loop ends below the kink) and 
no unit of area is added. Note that a unit of area may be added for 
any given output configuration. In this case the area generating
function is obtained correctly to $2W_{\rm max}+3k$. The factor $3k$ 
arise since it takes 3 lattice cells to completely surround the simplest 
puncture, which is just a single cell.   

As far as we are aware, series for punctured polygons have not been 
derived previously. In such circumstances it is even more important
than usual to undertake careful tests to ensure that the series are
correct. To this end, a second algorithm was implemented, by one of the 
authors, to independently evaluate the series coefficients. The complete
agreement we obtained between the two data sets reassures us as to
the correctness of our results.

\subsection{Enumeration of punctured staircase polygons 
\label{sec:stairenum}}

The enumeration of punctured staircase polygons is much simpler. In
fact as we shall demonstrate it is a problem for which the
computational complexity grows only as a polynomial in the
number of terms. As stated in the Introduction we can think
of staircase polygons as consisting of two non-intersecting
directed walks on the square lattice. Punctured staircase polygons
naturally will have more than two walks, in fact up to $2k+2$ walkers
can be present in any given column. Due to the restrictions on
staircase polygons it follows that a punctured staircase polygon   
is formed by requiring that the walks be directed and that
any two walkers starting at one point join each other later
without intersecting other walks. This can be encoded in a transfer
matrix calculation as follows: Again we count the number of
punctured polygons in rectangles as before and draw a line through
the lattice intersecting the $W+2$ edges. In this case we need
only specify whether or not each edge is part of the polygon or
not, so each configuration can be  represented by an ordered set of 
edge states $\{n_i\}$, where

\begin{equation*}
n_i  = \left\{ \begin{array}{rl}
 0 &\;\;\; \mbox{empty edge},  \\ 
1 &\;\;\; \mbox{part of loop closed to the left}.
\end{array} \right.
\end{equation*}
\noindent
This uniquely specifies the configuration because the first occupied 
edge is connected to the last occupied edge and any edges in between 
are paired, e.g., the second and third (fourth and fifth and so on) 
edges form a loop to the left and has to be connected to each other 
later on. The rules for updating the configurations are as follows: 
From the local input state `00' we can always get the output `00', and 
the output `11' provided there is an odd number of edges below the kink 
and also that the edge directly below the kink is empty. The rules for 
the output `11' ensure that the new pair of walkers lie within the 
enclosing staircase polygon but not inside an internal staircase 
polygon (thus preventing holes within holes), and that the lower walk 
does not intersect other walks. From the local input state `01' and `10' 
we can always produce the output `01', and the output `10' provided 
either that the edge directly below the kink is empty or there is an 
even number of edges below the kink. These rules ensure that the walkers 
do not intersect other walks except when we close a valid loop.
Finally from the local input state `11' we always get the
output state `00'. The weights associated with these updates
are obtained in the same way as for the SAP enumeration, whether
the enumeration is done by perimeter or area.

As in the previous case we calculate the number of punctured staircase 
polygons spanning the rectangles. Adding the contributions from all 
rectangles of width $W \leq W_{\rm max}$ and length 
$W \leq L \leq 2W_{\rm max}-W+1$ the number of punctured staircase polygons 
is obtained correctly up to perimeter $4W_{\rm max}+2+4k$. 
Note that for fixed $k$ the maximum number of configurations $N_C$ grows
as a polynomial in $W_{\rm max}$

\begin{equation*}
N_C = \sum_{j=0}^k  \left( \begin{array}{cc} 
W_{\rm max}+1 \\ 2j+2 \end{array} \right).
\end{equation*}
\noindent
So in this case the algorithm is of polynomial complexity.

\section{Expected behaviour}

As mentioned in the introduction, the problem of enumerating SAP by
area has been extensively studied \cite{EG90,Web}. It has been 
shown that $\kappa^{(k)} = \kappa^{(0)},$ and that if the exponents exist,
$\beta_k = \beta_0 + k$ \cite{WRen,Ren99}.  That is to say, for $k$ finite,
the connective constant for $k$-punctured discs,
by area, is the same as that for unpunctured discs, while the critical
exponent increases by 1 for each puncture. We repeat that the punctures
are disjoint.

As far as we are aware, there has been no previous work on the problem
of the perimeter generating function of punctured SAP. We first give a 
proof that a 1-punctured SAP on the square lattice has the same 
connective constant as an unpunctured SAP, and indicate how that proof can 
be generalised to $k$-punctured SAP.

Before stating and proving the relevant theorem, we require certain
preliminary results. For an (unpunctured) SAP of perimeter 
$2n$ one knows, \cite{Hug} eqn. (7.101), that 
\begin{equation}
 \exp[-b\sqrt{n}]\mu^{2n}/4n \le p_{2n}^{(0)} \le \mu^{2n}, 
\end{equation}
\noindent
where $b$ is a constant.
Further, it is obvious that a polygon of perimeter $2n$ has maximum area
$n^2/4,$ which occurs when the shape is an $n/2 \times n/2$ square.
Thus the area of a polygon of perimeter $2n,$ denoted $\A_{2n},$
satisfies $\A_{2n} \le n^2/4.$

Now consider 1-punctured polygons of total perimeter $2n,$
with inner perimeter $2m$ and outer
perimeter $2(n-m)$. It is possible for the inner polygon to be
Hamiltonian, in which case its perimeter is greater than that
of the surrounding polygon. Indeed, a square polygon of side
$2n + 1,$ and hence of perimeter $8n + 4$ can contain an internal polygon
of perimeter as large as $4n^2.$ Hence the semi-perimeter of the inner 
polygon, $m,$ can range from a minimum value of 2 to
a maximum value of $n - 2\sqrt{2n} + 2 < n - \sqrt{n}$ for $n > 1.$

With these preliminaries, we can now state and prove the following theorem:
\newline
{\bf Theorem:} $\lim_{n\rightarrow\infty}\frac{1}{2n}\log p^{(1)}_{2n} 
=\log \mu,$
where $\mu$ is the same constant as appears in the corresponding limit
for unpunctured SAP.
\newline
\noindent
{\bf Proof:}
1-punctured polygons of total perimeter $2n$ are constructed by placing
polygons $P$ of perimeter $2m$ inside polygons $Q$ of perimeter $2n-2m.$
Let $w(P,Q)$ denote the number of ways of placing polygon $P$ inside
polygon $Q,$ (which of course depends on both $P$ and $Q.$) Then
$$p_{2n}^{(1)} = \sum_{m=2}^{n-\sqrt{n}} \sum_{P} \sum_{Q} w(P,Q). $$
We bound this summand by the product of three factors. The first two
factors are the number of polygons of perimeter $2m$ and $2n-2m$
respectively. 
The third is the number of ways the smaller polygon can be 
placed inside the surrounding polygon and is clearly less 
than or equal to the area of the surrounding polygon. Explicitly,
$$p^{(1)}_{2n} \le \sum_{m=2}^{n - \sqrt{n}} (n-m)^2 \mu^{2m} \mu^{2(n-m)}/4  
\le  \mu^{2n}/4 \sum_{m=2}^{n} (n-m)^2 \le n^3\mu^{2n}.$$

To obtain a lower bound, consider
a $3 \times 3$ square polygon with a unit square hole at its centre.
This unique realisation of $p_{16}^{(1)}$ can be uniquely concatenated
with each unpunctured polygon by joining them at a solitary
specified edge, and
then deleting that edge. For each unpunctured polygon we take the 
set of left-most vertical edges and choose the bottom edge from this set, and
we choose the right-most, top-most vertical edge of $p_{16}^{(1)}$
(a similar operation is shown for polyominoes in figure~\ref{fig:concat}).
The concatenation operation then gives, for each of the $p_{2n-14}$
unpunctured polygons, a unique member of the set of 1-punctured polygons in 
which the puncture is a single cell. Thus we have
$p_{2n-14}^{(0)} \le p_{2n}^{(1)}.$
From the above equations we thus obtain
$$ \exp[-b\sqrt{n-7}]\mu^{2n-14}/4(n-7) \le p_{2n-14}^{(0)} 
\le p_{2n}^{(1)} \le n^3\mu^{2n} $$ for $n > 8.$
The theorem
then follows immediately on taking logarithms, dividing through
by $2n$ and taking the limit as $n \rightarrow \infty.$

This proof can clearly be extended to two-punctured
polygons, then to three-punctured polygons etc., by concatenating
unpunctured polygons with minimal two-punctured, three-punctured etc.
polygons. In the appendix we prove that $k$-punctured
polyominoes have the same growth constant as unpunctured SAP by area.

We have been unable to prove a result analogous to the result for the 
exponent of punctured SAP by area, but give an argument that depends
on certain assumptions that are generally accepted, though not proved.
In the case of staircase polygons however our assumptions have been proved,
and so our result will be rigorously true.

The key results we need are those obtained in \cite{EG90} to
the effect that the mean area of SAP of perimeter $2n$
is proportional to $n^{1.5}.$  This is true both
for SAP and for staircase polygons, and in the latter case it has been
proved. More precisely, we need the following result.
There exists constants $D_1$ and $D_2$ such that the mean area $\bar{A}_{2n}$
of polygons of perimeter $2n$ satisfies
$$D_1n^{\frac{3}{2}} \le \bar{A}_{2n} = \frac{\sum_Q A_Q}{p_{2n}}
\le D_2n^{\frac{3}{2}}, $$
where the sum is taken over all $p_{2n}$ polygons $Q$ of perimeter $2n.$

Further, there exists constants $C_1$ and $C_2$ such that the number
of polygons of perimeter $2n$ satisfies
$$C_1 \mu^{2n} n^{-\frac{5}{2}} \le p_{2n}^{(0)} 
\le C_2 \mu^{2n} n^{-\frac{5}{2}}.$$
For staircase polygons, $\mu = 2$ and the exponent is $\frac{3}{2}$ instead
of  $\frac{5}{2}$ in the above equation.

As above, let $Q$  denote a polygon of perimeter $2n-2m,$ with area $A_Q.$
The number of ways
of placing a given polygon $P$ of perimeter $2m$ inside $Q$
is clearly less than $A_Q.$

Thus the number of ways of placing $P$ inside any
polygon of perimeter $2n-2m$ is less than
\begin{equation}
\sum_{Q} A_{Q} = p_{2n-2m}^{(0)} \bar{A}_{2n-2m},
\end{equation}
where the sum is over all polygons $Q$ of perimeter $2n-2m.$

Hence the number of ways of placing all of the $p_{2m}^{(0)}$ polygons
of perimeter $2m$ inside any
polygon of perimeter $2n-2m$ is less than
$ p_{2m}^{(0)}p_{2n-2m}^{(0)}\bar{A}_{2n-2m},$ and so
\begin{equation}\label{eq:1pUB}
p_{2n}^{(1)} < \sum_{m=2}^{n-\sqrt{n}}  p_{2m}^{(0)}p_{2n-2m}^{(0)}
\bar{A}_{2n-2m} < D_2C_2^2\mu^{2n}2^{-5}S_{2n},
\end{equation}
where $$S_{2n} = \sum_{m=2}^{n-\sqrt{n}}m^{-\frac{5}{2}}/(n-m).$$

The last sum may be evaluated in a variety of ways. Using Maple we find
that $$S_{2n} = 0.341../n + O(n^{-\frac{3}{2}}).$$

Thus we obtain the bound
\begin{equation}
p_{2n}^{(1)} < const. \mu^{2n}/n.
\end{equation}
For staircase polygons the analogous calculation is slightly simpler,
and we find 
$$S_{2n} \sim 1+\zeta(3/2) +  O(n^{-\frac{1}{2}}),$$
where $\zeta(z)$ is the Riemann zeta function.

To obtain a lower bound, we restrict the inner polygon to be of perimeter 4, 
that is, a unit square. The number of polygons punctured by a unit square 
clearly provides a lower bound to the number of 1-punctured polygons.
A unit square can be placed anywhere in a polygon $Q$ of perimeter $2n - 4$
except on a boundary site. There are $2n - 4$ boundary sites. The mean area 
of $Q$ is greater than $  D_1(n-2)^{\frac{3}{2}},$ so we obtain the bound
\begin{equation}
p_{2n}^{(1)} >  p_{2n-4}(D_1(n-2)^{\frac{3}{2}} - 2n + 4) >
C_1\frac{\mu^{2n-4}}{(2n-4)^{\frac{5}{2}}}(D_1(n-2)^{\frac{3}{2}} - 2n)
> const. \mu^{2n}/n.
\end{equation}
Combining the two bounds gives the result
$$E_1 \mu^{2n}/n \le p_{2n}^{(1)} \le E_2 \mu^{2n}/n,$$
where $E_1 < E_2$ are constants. Accepting the usual asymptotic form
that is expected for such models, we conclude that
\begin{equation}
p_{2n}^{(1)} \sim const. \mu^{2n}/n.
\end{equation}
\noindent
(For staircase polygons the analogous result is 
$p_{2n}^{(1)} \sim const.4^n.$) 
Since the number of unpunctured polygons grows like
$\mu^nn^{-5/2},$ we see that 
the exponent is predicted to increase by $\frac{3}{2}$
as the result of a single puncture, while in the case of the
area generating function, the exponent is found to increase only
by 1. We show in the next section that this prediction is borne out
by our numerical calculations.

For punctured staircase polygons, a similar conclusion holds. That is,
the connective constant is unaltered at $\mu_{stair} = 2,$ but the exponent
increases by $1.5$ over its unpunctured counterpart
when enumerating 1-punctured staircase polygons by perimeter.
In fact, we have been able to calculate the generating function for
staircase polygons with a single puncture of perimeter 4, and also with
a single puncture of perimeter 6. The generating functions for these
special cases are given below, and
are precisely in accordance with the more general results
given above.

We also note that Cardy \cite{Ca99} recently considered the problem
of the number of punctured SAP with $k$ concentric, mutually self-avoiding
SAPs surrounding a fixed point of the dual lattice. Thus for $k = 1$ this
corresponds to 1-punctured SAPs surrounding a fixed point. In that case
Cardy finds \cite{Ca99} for the number of such configurations $b_{2n}^{(1)}$
that $b_{2n}^{(1)} = \mu^{2n}\frac{\ln{n}}{64\pi^2 n}.$ The above calculation 
may be repeated for Cardy's problem. All that is required is to add a factor 
$\bar{A}_m$ to the summand in eqn.~\ref{eq:1pUB}, as the
surrounded point can be anywhere inside the inner polygon. Thus we must
multiply by the mean area of that polygon. The sum defining $S_{2n}$ then
becomes $S_{2n} = \sum_{m=2}^{n-\sqrt{n}} \frac{1}{m(n-m)} \sim \log{n}/n,$
in agreement with Cardy's result. (Our lower bound for this problem is too
weak, but with more effort could be improved.)

\section{Analysis of the series \label{sec:analysis}}

All the series we have investigated are characterised by coefficients
that grow exponentially, with sub-dominant term given by a
critical exponent. The generic generating function behaviour
is $G(z) =\sum_n g_n z^n \sim A(z)(1 - \sigma z)^{-\xi},$ 
and hence the coefficients
of the generating function $g_n = [z^n]G(z) \sim A(1/\sigma)/\Gamma(\xi) 
\sigma^n n^{\xi-1}.$
Generally speaking the existence of the growth constant $\sigma$ has
been proved, but except for exactly solvable models, such as staircase
polygons, the existence of the critical exponent $\xi$ has only
been conjectured, though its existence has never been doubted.
The radius of convergence of the generating
function is usually given by the critical point, which is
at $z = 1/\sigma.$

We principally used two methods to analyse all the series studied
in this paper. Firstly, to obtain the singularity structure of
the generating function we used  the
numerical method of differential approximants \cite{Guttmann89}.
In particular, we used this method to estimate the growth
constant $\sigma$ and the critical exponent $\xi.$ We were
invariably able to conjecture an exact value for $\xi,$ which
was always an integer or half-integer for all the problems we
investigated.  Imposing this conjectured exponent permitted a
 refinement of the estimate of the growth constant
--- providing so-called biased estimates.

Once the exact value of the exponent was conjectured, and the
growth constant accurately estimated, we turned our attention
to the ``fine structure'' of the asymptotic form of the
coefficients, by fitting the coefficients to the assumed form
$g_n = [z^n]G(z) \approx \sigma^n n^{\xi-1}\sum_{i\ge0}c_i/n^{f(i)}.$
If there is no non-analytic correction term, then $f(i) = i,$
while a square-root correction term means $f(i) = i/2.$ For
all the series studied, only these two situations were encountered.

In all cases, our procedure is to {\em assume} a particular
form for $f(i),$ and see how it fits the data. With the very long
series we now have at our disposal, it is usually easy to see
if the wrong assumption has been made --- the sequence of
amplitude estimates $c_i$ either diverges to infinity
or converges to zero.
Once the correct assumption is made, convergence is usually
rapid and obvious. A detailed demonstration of the method can
be found in \cite{CG96,JG99}.

As an example of the sort of results we obtained, we show 
sequences of estimates of the
coefficients of the perimeter generating function of unpunctured
SAP in Table~\ref{tab:fit}.
In that case we conjectured that $f(i) = i.$
Because of the large amount
of tabular data generated by the method, we have not given
this level of detail for the many series investigated
here. We show only the results for two series.
For the others,  we just give our assessment of the 
apparent convergence of the sequences $c_i,$ and the estimated
value of the limits. As the equations involved are linear, the
method is easy to implement, and interested readers can
readily generate the relevant data themselves.
Some subtleties nevertheless exist. For example, for
punctured staircase polygons, the perimeter generating function
has {\em two} singularities on the circle of convergence, and
so both must be taken into account. We discuss this in more detail
in the relevant section below.

As for the first stage of the analysis, the method of
differential approximants, we proceeded as follows:
Estimates of the critical point and critical exponent
were obtained by averaging values obtained from first order
$[L/N;M]$ and second order $[L/N;M;K]$ inhomogeneous differential
approximants. For each order $L$ of the inhomogeneous polynomial we 
averaged over those approximants to the series which used  at least 
the first 80\% - 90\% of the terms of the series,
and used approximants such that the difference between $N$, $M$, and $K$ 
didn't exceed 2. These are therefore ``diagonal'' approximants.
Some approximants were excluded from
the averages because the estimates were obviously spurious. The error 
quoted for these estimates reflects the spread (basically one standard
deviation) among the approximants. Note that these error bounds should
{\em not} be viewed as a measure of the true error as they cannot include
possible systematic sources of error. However systematic error
can also be taken into account in favourable situations, as
for example in the case of SAP enumerated by perimeter \cite{JG99}.
Again, in the interests of space, we present only our results,
and not the intermediate detail from which our estimates were made.
An example in full detail for one of the series investigated
in this study can be found in \cite{JG99}.
We turn now to the analysis of all the series.

\subsection{Staircase polygons}

For (unpunctured) staircase polygons, the multi-variable width, height
and area generating function is known \cite{BM96}. As usual, we
denote
$$(a)_n = \prod_{i=0}^{n-1} (1 - aq^i).$$ Further, denoting the first
two $q$-Bessel functions as:
$$ J_0(x,y,q) = \sum_{n\ge0}\frac{(-1)^nx^nq^{n+1 \choose 2}}{(q)_n(yq)_n},$$
and
$$ J_1(x,y,q) = \sum_{n\ge1}\frac{(-1)^{n-1}x^nq^{n+1 \choose 2}}
{(q)_{n-1}(yq)_{n-1}(1-yq^n)},$$
the perimeter and area generating function is simply
$$ P(x,y,q) = y\frac{J_1(x,y,q)}{J_0(x,y,q)},$$
where $x$ $(y)$ is the variable conjugate to the horizontal (vertical)
semi-perimeter,
while $q$ counts the area. No analogous results are known for punctured 
staircase polygons, though the calculation for SAP \cite{WRen} by area can 
be carried over {\em mutatis mutandis} to prove that the area generating 
function for a punctured staircase polygon with $k$ holes has the same radius
of convergence as the area generating function for unpunctured staircase
polygons. Further, the critical exponent increases by 1 for each puncture,
and unlike the case for SAP, we can not only prove the existence of a critical
exponent for unpunctured staircase polygons, but we know its value.
Hence we know the leading term in the asymptotic expansion of the
generating function by area for $k$-punctured staircase polygons.

For the expected behaviour of the perimeter generating function
of punctured staircase polygons, 
the arguments of the preceding section apply directly. The
radius of convergence, and hence the connective constant remains unchanged, 
and the argument given in the preceding section suggests that the
critical exponent should increase by 1.5 for each puncture. 
The results of our analysis, presented below, bear this out.

\subsubsection{Area generating function}

For unpunctured staircase polygons, the area generating function is given 
by $$ A(q) = \sum_{n\ge1} a_n^{(0)}q^n =\frac{J_1(1,1,q)}{J_0(1,1,q)}.$$

By inspection, this has poles at the zeros of $J_0(1,1,q).$
The nearest zero is at \cite{BE74} 
$1/q = \eta = 2.30913859330,$ 
and there is a simple pole at that point. The next zero is
well separated (at $1/q = \lambda = 1.4435..$) and so the asymptotic
form of the generating function is
$$ A(q) \sim D/(1 - \eta q) + E/(1 - \lambda q) + \ldots, $$
and hence
$$a^{(0)}_n = [q^n]A(q) \sim \eta^n(c_0 + O((\lambda/\eta))^n).$$
Our analysis bears this out, and we estimate $c_0 = 0.12881579.$

For 1-punctured discs, our analysis, based on more than 100 series
coefficients, convincingly suggests the following asymptotic form:
$$ A^{(1)}(q) \sim D^{(1)}/(1 - \eta q)^2 + E^{(1)}/(1 - \eta q)^{1.5} + 
F^{(1)}/(1 - \eta q) + \ldots, $$
and hence
$$a^{(1)}_n = [q^n]A^{(1)}(q) \sim \eta^n n\sum_{i\ge0}c_i/n^{i/2}.$$
The sequences of amplitude estimates, assuming this asymptotic form,
are shown in Table~\ref{tab:fit2}. The apparent convergence of the
amplitude estimates is, as explained above, our source of evidence
for this asymptotic form.

For 2-punctured discs, a similar analysis, based on some 86 series
coefficients, convincingly suggests the following asymptotic form:
$$ A^{(2)}(q) \sim D^{(2)}/(1 - \eta q)^3 + E^{(2)}/(1 - \eta q)^{2.5} 
+ \ldots, $$
and hence
$$a^{(2)}_n = [q^n]A^{(2)}(q) \sim \eta^n n^2\sum_{i\ge0}a_i/n^{i/2}.$$

For 3-punctured discs our analysis was based on an 89 term series.
We found that the above pattern persists, so that the generating
function has
the following asymptotic form:
$$ A^{(3)}(q) \sim D^{(3)}/(1 - \eta q)^4 + E^{(3)}/(1 - \eta q)^{3.5} 
+ \ldots, $$
and hence
$$a^{(3)}_n = [q^n]A^{(3)}(q) \sim \eta^n n^3\sum_{i\ge0}a_i/n^{i/2}.$$

Estimates of the various amplitudes defined above are shown in
Table~\ref{tab:amps}.

\subsubsection{Perimeter generating function}

For unpunctured staircase polygons, the perimeter generating function 
(ignoring the distinction between height and width) is given by
$$ P(x) = \frac{1 - 2x -\sqrt{1 - 4x}}{2},$$ which is, apart from 
suppression of the first term, the generating function for Catalan 
numbers. Hence
$$p^{(0)}_{2n} = [x^n]P(x) = \frac{1}{n} {{2n-2} \choose{n-1}} 
\sim 4^n/n^{\frac{3}{2}}\sum_{i\ge0}c_i/n^{i}.$$
The values of $c_i$ follow immediately from the exact solution. They
are given in Table~\ref{tab:amps}.

For 1-punctured discs, our analysis, again based on more than 100 series
coefficients, was more equivocal than that of the 1-punctured area generating
function. 
The method of differential approximants clearly identified a singularity
at the known critical point, $x_c = 1/4,$ but almost all approximants had
a double root, implying a confluent singularity. The leading exponent was
estimated to be $-1,$ implying a pole in the generating function, (and
hence immediately lending support to our conjectured change in the critical
exponent of 3/2 as a result of puncturing), but
we were unable, from this method, to identify the confluent exponent.
Further, a second singularity was identified, of the form
const.$(1 + x/x_c)^{6.5}.$
At this point we wish to remark on the close similarity
between a recently solved model of polygons, the {\em three-choice
polygon} model \cite{CDG} and 1-punctured staircase polygons.
1-punctured staircase polygons can
be thought of as being constructed from two three-choice polygons, with
common edges deleted. This geometric similarity is borne out by the
fact that the two models have identical singularity distributions,
with even the exponents being the same at all non-physical singularities.

For 1-punctured discs then, the perimeter generating function is 
expected to be of
the following asymptotic form:
$$ P^{(1)}(x) =  \sum_{n} p^{(1)}_{2n} x^{n} \sim B^{(1)}(x) 
+ C^{(1)}(x)(1 - 4x)^{-1} + D^{(1)}(x)(1 + 4x)^{6.5},$$
and hence
$$p^{(1)}_n = [x^n]P^{(1)}(x) \sim 4^n \sum_{i\ge0}c_i/n^{f(i)}
+ (-4)^n n^{-7.5}\sum_{i\ge0}d_i/n^{i},$$
where $B^{(1)}(x),$ $C^{(1)}(x)$ and $D^{(1)}(x)$ are assumed regular in
the disc $|4x| \le 1.$

Assuming $f(i) = i,$ which implies only analytic correction-to-scaling
terms, gave unsatisfactory results. Notably, we observed that the estimates 
of the amplitude $c_1$ in the above asymptotic form were steadily increasing,
suggesting that the assumed form did not properly account for the 
correction-to-scaling terms. With $f(i) = i/2$ the amplitude estimates were 
much more stable. This then implies that the generating function in fact 
behaves as
\begin{equation}
 P^{(1)}(x) \sim B^{(1)}(x) + C^{(1)}(x)(1 - 4x)^{-1} 
 + E^{(1)}(x)(1 - 4x)^{-\frac{1}{2}}
 + D^{(1)}(x)(1 + 4x)^{6.5}. 
\end{equation}
The amplitude estimates, assuming this asymptotic form,
are shown in Table~\ref{tab:amps}. The apparent convergence of the
amplitude estimates is, as explained above, our source of evidence
for this asymptotic form. We did not tabulate the non-physical amplitudes,
as they are of little interest to our investigation. However they need
to be included to stabilise estimates of the physical amplitudes. We
mention in passing that $d_0 \approx 0.14.$

We also generated series for staircase polygons with two and
three punctures, with perimeter 150 and 134 steps respectively.
Our differential approximant analysis lent support to our expectation
that the critical exponent increases by $1.5$ per puncture. This turned
out to be true for the non-physical singularity at $x = -\frac{1}{4}$
as well.
A similar analysis to that for the one-punctured
polygons strongly supported the analogous asymptotic forms,
\begin{equation}
P^{(k)}(x)  =  \sum_{n} p^{(k)}_{2n} x^{n} \sim B^{(k)}(x) + 
C^{(k)}(x)(1 - 4x)^{0.5 - 1.5k} + D^{(1)}(x)(1 + 4x)^{8 - 1.5k}
\end{equation}
for $k > 0.$
\noindent
Hence the asymptotic form of the coefficients is conjectured to be
$$ p^{(k)}_{2n} \sim 4^{n}n^{\frac{3(k-1)}{2}}
\sum_{i\ge0}c_{i}^{(k)}/n^{i/2} + (-4)^n n^{1.5k-9}
\sum_{i\ge0}d_{i}^{(k)}/n^{i},$$
where the amplitude estimates $c_{i}^{(k)}$ are given in Table~\ref{tab:amps}.

We have also determined the exact generating function for staircase
polygons punctured by a single hole of perimeter 4, and also the generating
function for staircase polygons punctured by a single hole of perimeter
6. We obtained these generating functions by generating the
coefficients using the algorithm discussed, and then searching for
an underlying differential equation. As a result we find the perimeter
generating functions $P_4^s(x)$ and $P_6^s(x)$
for 1-punctured staircase polygons with a hole of perimeter 4 and 6 respectively.

\begin{equation}
P_4^s(x)= \frac{2x^4  - 16x^3  + 20x^2  - 8x + 1}{2(1 - 4x)}
 - \frac{1 - 6x + 10x^2  - 4x^3 }{2\sqrt{1 - 4 x}}
\end{equation}
and
\begin{eqnarray}
      P_6^s(x) &=&  
\frac{1-26x+228x^2-906x^3+1709x^4-1378x^5+322x^6 }{2(1 - 4x)^{\frac{5}{2}}}
    \nonumber \\
    & & -\frac{32x^6-404x^5+815x^4-586x^3 +182x^2-24x+1}{2(1 - 4x)^2}.
\end{eqnarray}

\noindent
Note that both these exact solutions display the confluent square root 
correction that we have found in our numerical investigations in the more
general case.
In the next subsection we analyse the analogous generating function for
SAP.

\subsection{Self-avoiding polygons \label{sec:SAPana}}

For unpunctured SAP, the perimeter generating function was recently
extended \cite{JG99} to 90 step polygons, and the asymptotics clearly 
identified. The polygon generating function is defined to be
\begin{equation}
P^{(0)}(x)  =  \sum_{n} p_{2n}^{(0)} x^{n} \sim A^{(0)}(x) + B^{(0)}(x)
(1-\mu^2 x)^{\frac{3}{2}},
\end{equation}
\noindent
where the functions $A^{(0)}$ and $B^{(0)}$ are believed to be regular in 
the vicinity of $x_c=1/\mu^2.$ We estimated $\mu =2.63815853034(10)$.
From this equation follows the asymptotic form of the coefficients,
\begin{equation}
p_{2n}^{(0)} \sim \mu^{2n} n^{-\frac{5}{2}}[c_1 + c_2/n + c_3/n^2 
+ c_4/n^3 + \cdots].
\end{equation}
We show in Table~\ref{tab:fit} the sequence of estimates of ${c_i},$
and in Table~\ref{tab:amps} our estimates of the amplitudes,
being the limits of the sequences $\{c_i\}.$

The area generating function was first studied in \cite{EG90}, where
the first 20 terms were given, and
the asymptotic form estimated to be
\begin{equation}
A^{(0)}(q)  =  \sum_{n} a_{n} q^{n} \sim D(q) + E(q)\log (1- \kappa q),
\end{equation}
\noindent
where $\kappa \approx 3.97087,$ and the logarithm in the above equation
was understood to include the possibility of some power of a logarithm 
other than unity. (Though our analysis below implies that this is not the 
case.) In the present work we extend the series to 42 terms.

\subsubsection{Area generating function}

Using our greatly extended 42 term series, our analysis of the
unpunctured disc area generating function was carried out by standard 
methods. We used differential approximants \cite{Guttmann89} and found
unbiased critical point and critical exponent estimates. The unbiased
exponent estimate had absolute value less than $10^{-5},$ totally
supporting our view that it is exactly zero. Assuming this, a biased
estimate of the critical point is possible, and in this way we estimate
$\kappa = 3.97094397(9).$

We then proceeded to seek the asymptotic form of the coefficients
by writing
$$a_n = [q^n]A^{(0)}(q) \sim \kappa^n/n \sum_{i\ge0}c_i/n^{f(i)}.$$

Our numerical results were well converged, demonstrating very convincingly 
that $f(i) = i.$ This is the asymptotic form consistent with a pure
logarithmic singularity, not raised to any power. Estimates of the
amplitudes are given in  Table~\ref{tab:amps}.

For 1-punctured discs, our analysis, based on the series known to $q^{45}$,
convincingly suggests the following asymptotic form:
$$ A^{(1)}(q) \sim D^{(1)}(q) + E^{(1)}(q)/(1 - \kappa q), $$
and hence
$$a^{(1)}_n = [q^n]A^{(1)}(q) \sim \kappa^n \sum_{i\ge0}c_i/n^{i}.$$
Attempts to fit to alternative forms, corresponding to a confluent logarithm
or a non-analytic correction-to-scaling term were unsuccessful, adding
to our confidence that the above form is correct.
Estimates of the amplitudes are given in  Table~\ref{tab:amps}.

For twice-punctured discs, our analysis, also based on the series known
to $q^{48}$ convincingly suggests the following asymptotic form:
$$ A^{(2)}(q) \sim D^{(2)}(q) + E^{(2)}(q)/(1 - \kappa q)^2, $$
and hence
$$a^{(2)}_n = [q^n]A^{(2)}(q) \sim \kappa^n n\sum_{i\ge0}c_i/n^{i}.$$

The conjectured asymptotic form for $k$ punctured polygons, by area,
is thus
$$ A^{(k)}(q) \sim D^{(k)}(q) + E^{(k)}(q)/(1 - \kappa q)^k. $$

\subsubsection{Perimeter generating function}

The results for unpunctured polygons are fully discussed at 
the beginning of subsection \ref{sec:SAPana}.
As we found with staircase polygons, the generating function for
punctured discs by perimeter was a more challenging numerical
analysis problem than either its unpunctured counterpart, or its
area counterpart.

We found that the method of differential approximants was not
particularly satisfactory. Given that we needed some 100 terms
to successfully analyse the (presumably simpler) problem of
punctured staircase polygons by perimeter, it is not surprising
that for punctured SAP, for which we have 33 non-zero coefficients
(corresponding to perimeters up to 84 steps), the method was not
satisfactory. However, it did indicate the presence of a confluent
singularity. As we found a confluent square-root singularity for
staircase polygons, it is hardly surprising that a confluent singularity
is detected for the punctured SAP generating function.
In fact an
exponent shift of around 1.7 was seen, compared to the expected value
1.5. We attribute this to the ``short'' series, coupled with
the well known deleterious effect of confluent terms
in such an analysis.
Nevertheless, subsequent analysis of the asymptotic form of
the coefficients, assuming an exponent shift of 1.5, together
with a square-root confluent term, as found for punctured staircase
polygons, gave satisfactory results.

We denote the generating function for $k$-punctured SAP, by perimeter,
as
\begin{equation}
P^{(k)}(x)  =  \sum_{n} p^{(k)}_{2n} x^{n} \sim B^{(k)}(x) 
+ C^{(k)}(x)(1- \mu^2 x)^{1.5 - 1.5k},
\end{equation}
where the exponent is conjectured.
\noindent
For 1-punctured SAP, the vanishing of the exponent implies a logarithmic
singularity. We fitted the coefficients to the asymptotic form appropriate
to $\log(1- \mu^2 x),$
so that the asymptotic form of the
coefficients just involves decreasing integer powers of $n.$
We then found that the estimates of the leading amplitude were
monotonically increasing, which implies that the asymptotic form is wrong ---
too weak. Including a confluent square root singularity, as was found
for punctured staircase polygons, stabilised the estimates.
Accordingly, we conjecture that
the asymptotic form is dominated by a logarithmic singularity, with 
a sub-dominant square root singularity, so that
\begin{equation}
p^{(1)}_{2n} = [x^n]P^{(1)}(x) \sim \mu^{2n}/n\sum_{i\ge0}c_i/n^{\frac{i}{2}}
\end{equation}
Estimates of the amplitudes $c_i$ are given 
 in Table~\ref{tab:amps}.

For twice punctured discs, a similar analysis suggested that the
asymptotic form of the generating function is
\begin{equation}
P^{(2)}(x)  =  \sum_{n} p^{(2)}_{2n} x^{n} \sim B^{(2)}(x) 
+ C^{(2)}(x)(1-\mu^2 x)^{-1.5},
\end{equation}
\noindent
again with evidence of a square root confluent term. As for 1-punctured SAP, 
 we give estimates of the amplitudes $c_i$ defined by
$$ p^{(2)}_{2n} \sim \mu^{2n}n^{\frac{1}{2}}
              \sum_{i\ge0}c_{i}/n^{\frac{i}{2}} $$
in  Table~\ref{tab:amps}.

For thrice punctured discs, a similar analysis suggested that the
asymptotic form of the generating function is
\begin{equation}
P^{(3)}(x)  =  \sum_{n} p^{(3)}_{2n} x^{n} \sim B^{(3)}(x) 
+ C^{(3)}(x)(1-\mu^2 x)^{-3},
\end{equation}
\noindent
again with evidence of a square root confluent term.
As for 1-punctured SAP, 
 we give estimates of the amplitudes $c_i$ defined by
$$ p^{(3)}_{2n} \sim \mu^{2n}n^2\sum_{i\ge0}c_{i}/n^{\frac{i}{2}} $$
in  Table~\ref{tab:amps}.

The conjectured asymptotic form for $k$ punctured polygons, by perimeter,
is then
\begin{equation}
P^{(k)}(x)  =  \sum_{n} p^{(k)}_{2n} x^{n} \sim B^{(k)}(x) 
+ C^{(k)}(x)(1-\mu^2 x)^{1.5 - 1.5k}
\end{equation}
where for $k > 0$ we find strong evidence for a square root 
correction-to-scaling term.

\subsection{Polyominoes}

The problem of polyominoes has a long and interesting history, and has
been well discussed in the popular scientific literature \cite{Gol}.

The enumeration of square lattice polyominoes to 24 steps \cite{RE81}
was given in 1981, extended to 25 steps in 1995 \cite{CG95} and
currently stands at 28 steps \cite{Web}.

We have analysed the latest series by the method of differential 
approximants, and find the generating function behaves as
\begin{equation}
\PP(y)  =  \sum_{n} \mathit{a}_{n} y^{n} \sim G(y) + H(y)\log (1- \tau y),
\end{equation}
\noindent
where $a_n$ is the number of polyominoes of area $n$.
In \cite{CG95} the estimate $\tau = 4.06265(5)$ was given. The extra terms
now available allow us to make the refined estimate $\tau = 4.062591(9).$
Analysis of the asymptotic form of the coefficients is totally consistent
with a simple logarithm in the generating function. Thus
$$\mathit{a}_n = [y^n]\PP(y) \sim \tau^n \sum_{i\ge0}c_i/n^{i+1}.$$
The amplitude estimates are given in Table~\ref{tab:amps}. The leading
amplitude is in complete agreement with, but 3 orders of magnitude more
accurate than that given in \cite{GU82}, while the order of the leading
term --- $O(1/n)$ --- was predicted by physical arguments back in 1981
\cite{PS81}. In that work, the logarithmic singularity in the generating
function of strongly embedded site-animals was obtained. As discussed in
the introduction, these are isomorphic to polyominoes.

As well as extending the polyomino series, Oliveira e Silva has enumerated
$k$-punctured polyominoes \cite{Web} for $k \le 6.$ Clearly, $k = 0$ 
polyominoes are just SAP, and as we have seen in the previous section, 
these grow as $\kappa^n$ where $\kappa = 3.9709.. < \tau.$ The arguments 
in \cite{Ren,WRen} can be modified and applied to
show that a $k$-punctured polyomino has the same growth constant $\lambda$
as its unpunctured counterpart, and this is done in the Appendix.

Thus the situation is that, for any finite number of punctures,
$$\mathit{a}_n^{(k)} = [y^n]\PP^{(k)}(y) \sim \kappa^n ,$$
but that
$$\mathit{a}_n = [y^n]\PP(y) = \sum_{k\ge0} \mathit{a}_n^{(k)}
\sim \tau^n .$$

We have analysed the series for $k$-punctured polyominoes, $ k = 0, 1, 2,$
and find the asymptotic form of the coefficients to be

$$\mathit{a}_n^{(k)} = [y^n]\PP^{(k)}(y) \sim \kappa^n
n^{k-1}\sum_{i\ge0}c_i^{(k)}/n^{i},$$
corresponding to a generating function for $k$-punctured polyominoes
having a $k^{th}$ order pole, viz:
\begin{equation}
\PP^{(k)}(y)  =  \sum_{n} \mathit{a}_{n}^{(k)} y^{n} \sim
H^{(k)}(y)(1- \kappa y)^{-k},
\end{equation}
\noindent
where $k=0$ is to be interpreted as a logarithm.
Thus just as for punctured polygons, it is found that the exponent of
$k$-punctured polyominoes increases by 1 for each puncture. This is also
proved in the Appendix.

The coefficients $c_i^{(k)}$ are given in  Table~\ref{tab:amps}.

\section{Conclusion}

We have investigated the effect of punctures on SAP and staircase 
polygons enumerated both by area and perimeter.

In order to do this, we have developed a new algorithm, exponentially
faster than direct counting, whereby we have radically extended a number of 
series. This extension was necessary in order to probe some rather subtle
numerical behaviour.

We found that, in every case, a finite number of punctures does not
change the exponential growth factor associated with the unpunctured
counterpart of the punctured object being enumerated.

This latter conclusion was also proved for polyominoes. The
effect of punctures was also investigated numerically for polyominoes.

Writing $b_n^{(k)}$ for the $n^{th}$ coefficient in the generating 
function for some $k-$punctured object, so that
$$b_n^{(k)} \sim (\omega^{(k)})^n n^{\theta(k)}, $$
we found $ \omega^{(k)} =  \omega^{(0)}$ in all cases.
We found, further, that  $\theta(k) = \theta(0) + k$
if enumerating any of the objects we have considered by area. This
can be proved, though subject, in some cases, to the existence of
the exponent in question. Subject to certain unproved assumptions
we  also showed that
$\theta(k) = \theta(0) + 3k/2$
if enumerating by perimeter.

We have, for the first time, obtained good numerical estimates of 
the sub-dominant terms for a range of problems, thus identifying both
the nature of the generating function and any correction-to-scaling terms.

We have also obtained an exact solution for the generating functions
of staircase polygons, enumerated by perimeter, punctured by
a single hole of perimeter 4 and of perimeter 6. These exact
solutions provide additional support for our numerically based
conjectures of the correction-to-scaling exponent in the general case.

A more accurate estimate of the growth constant for SAP
enumerated by area has been given, complementing our earlier work
on the perimeter growth constants \cite{JG99}.

\section*{E-mail or WWW retrieval of series}

The series for the various generating functions studied in this paper
can be obtained via e-mail by sending a request to 
I.Jensen@ms.unimelb.edu.au or via the world wide web on the URL
http://www.ms.unimelb.edu.au/\~{ }iwan/ by following the instructions.

\section*{Acknowledgements}

We have derived great benefit from discussions of aspects of this problem 
with Mireille Bousquet-M\'{e}lou, John Cardy, Aleks Owczarek, 
Buks van Rensburg, and  Stuart Whittington. We are particularly
grateful to Mireille Bousquet-M\'{e}lou for her careful reading of the
manuscript, to Buks van Rensburg for clarifying aspects of his earlier 
work, and to John Cardy for querying some earlier, incorrect results.
IJ and AJG gratefully acknowledge financial support from the 
Australian Research Council. LHW would like to thank 
The University of Melbourne for the Melbourne Research Scholarship.

\setcounter{section}{1}
\setcounter{subsection}{0}
\setcounter{subsubsection}{0}
\setcounter{equation}{0}
\def\thesection{\Alph{section}}
\def\theequation{A\arabic{equation}}

\newtheorem{theoremapp}{Theorem}
\newtheorem{lemmaapp}{Lemma}
\newtheorem{corollaryapp}{Corollary}
\def\thetheoremapp{A\arabic{theoremapp}}
\def\thelemmaapp{A\arabic{lemmaapp}}
\def\thecorollaryapp{A\arabic{corollaryapp}}

\section*{Appendix. Growth constants and exponents for $k$-punctured 
polyominoes}

In this appendix we show that the growth constants are the same for 
all $k$-punctured polyominoes. Our method of proof is based on that of 
van Rensburg and Whittington \cite{Ren,WRen}, making the necessary 
changes for the polyomino problem, and discussing in detail certain 
special cases. That is to say, if $\kappa$ denotes the connective 
constant for SAP enumerated by area, then this is the  growth constant 
for $k$-punctured polyominoes for any  finite $k$.  Further,  if the 
usual asymptotic form for the number of $k$-punctured polyominoes is 
assumed, $s_n^{(k)} \sim C_k n^{-\phi_k}\kappa^n$, and $\phi_0$ exists,
then  $\phi_k = \phi_0 - k.$

\subsection{Operations and mappings on punctured polyominoes}

Let the set of all $k$-punctured polyominoes with $n$ cells be denoted by 
$\Phi_{n}^{(k)}$, and the set of all polyominoes of $n$ cells be $\Phi_n$. 
Then, for each $n$, $\Phi_{n} = \bigcup_k {\Phi_{n}^{(k)}}$.  Throughout, 
let $s_n^{(k)}$ denote the cardinality of $\Phi_{n}^{(k)}$ and let $s_n$ 
denote the cardinality of $\Phi_{n}$. Hence $s_n = \sum_k {s_n^{(k)}}$. 

Following \cite{Ren, WRen} we now define operations on punctured polyominoes, 
including concatenation, drilling and surgery, which are needed in subsequent 
proofs. Concatenation allows us to change the size of polyominoes, while 
drilling and surgery are concerned with changing the number of holes in a 
polyomino.

\subsubsection{Concatenation \label{sec:concat}}

The concatenation mapping defined here is similar to that in \cite{Ham61}.
Consider the bounding rectangle $R(P)$ of any polyomino 
$P \in \Phi_n^{(k)}$. Define the top (bottom) edge of 
$P$ to be the top (bottom) edge along the east (west) side of 
$R(P)$.

Now, the concatenation of two polyominoes $P \in \Phi_n^{(h)}$ 
and $Q \in \Phi_m^{(k)}$ is defined by joining $P$ 
and $Q$ while superimposing the top edge of $P$ 
and the bottom edge of $Q$.  The result is an $(h+k)$-punctured 
polyomino with $m+n$ cells,  see figure~\ref{fig:concat}.  
Hence we have a map

\begin{equation}
T: \Phi_n^{(h)} \times \Phi_m^{(k)} \longmapsto \Phi_{m+n}^{(h+k)} 
\end{equation}

\begin{lemmaapp} \label{lem:incmono}
For all non-negative values of $h$ and $k$, $s_{n}^{(h)}s_m^{(k)} 
\le s_{m+n}^{(h+k)}$ and in particular, $s_{n+1}^{(h)} \ge s_n^{(h)}$
\end{lemmaapp}
\textit{Proof:} For the mapping $T$ defined above, every pair of polyominoes 
in the domain can be concatenated to form a larger polyomino in the codomain. 
Conversely, every such polyomino in the codomain can be uniquely broken up 
into the original ones. However, there are some polyominoes in the codomain 
which cannot be formed by concatenating two smaller polyominoes.  An example 
is a $2 \times 2$ polyomino. Hence we get the first part of lemma as 
\[
|\Phi_n^{(h)} \times \Phi_m^{(k)}| \le |\Phi_{m+n}^{(h+k)}|.
\]  
Putting $k=0$ and $m=1$ and noting that $|\Phi_1^{(0)}| = 1$, we get the 
second part. 
$\quad \blacksquare$

\subsubsection{Drilling \label{sec:drill}}

Simply creating a hole inside an unpunctured polyomino by removing some 
interior cells does not allow us to drill certain classes of polyominoes.  
For instance, a polyomino composed of a single linear sequence of cells 
cannot be `drilled' since  removing any cell will either disconnect the
polyomino or shorten the sequence.  

The following definition of drilling differs somewhat from that in 
\cite{Ren, WRen}, though the underlying idea is the same.

First, we drill one hole.  Let $P_0 \in \Phi_n^{(0)}$ be the
polyomino that we are to puncture.  Cover $P_0$ by a grid system, 
with each grid square of size $b \times b$ cells for any $b \ge 5$ (the 
minimum size will be justified later).  Say $b=5.$  Pick any grid square 
$G$ that covers at least one cell of $P_0$ and drill a hole there, 
as detailed below.

\textbf{Step 1} Remove all cells within the selected grid square $G$.  If, 
after this step, we are left with a $1$-punctured polyomino, we are done.  
If not, go to step 2.

\textbf{Step 2} Check the corners of $G$:  If there are 2 disconnected 
components touching each other only at the corner, connect them by adding a 
cell at the appropriate corner of $G$.

\textbf{Step 3} Put a $1$-punctured polyomino with 8 cells at 
the centre of $G$.

\textbf{Step 4} Reconnect the disconnected components outside $G$ to the 
$1$-punctured polyomino by adding linear sequences of cells (non-unique).  
These steps are illustrated in figure~\ref{fig:drill}.

Note that the minimum value of $b$ is 5 because if we have anything less 
than that, we might create extra holes unexpectedly as the $1$-punctured 
polyomino in \textbf{Step 3} must touch the boundary of the grid square.  
An example in figure~\ref{fig:exdrill}  will illustrate this.  In this 
example the original polyomino has no holes and the drilled polyomino has 
two holes, whereas with a grid of size 5 (or more), the number 
of holes in the drilled polyomino is only one.

After this operation, the maximum number of cells that could be removed is 
$b^2 = 25$ (finishing at step 1).  The maximum number of cells that could be 
added is $b^2 - 2$ (this occurs when there was only 1 cell in the square 
before drilling and we end up with a 1-punctured polyomino with $b^2 - 1$ 
cells after the operation).  So, depending on each instance of the drilling 
operation, we obtain a resulting polyomino $P_1 \in \Phi_j^{(1)}$ 
where $j$ could be anything  from $n-b^2$ to $n+(b^2-2)$.  The drilling 
operation thus defines a map

\begin{equation}
D: \Phi_n^{(0)} \longmapsto \bigcup_{j = n - b^2}^{n + (b^2 -2)} \Phi_j^{(1)}.
\end{equation}

\begin{theoremapp}
There exists a real constant $C$ such that for all $b \ge 5$,
\begin{equation}
s_n^{(0)} \le C s_{n+(b^2 - 2)}^{(1)}.
\end{equation} 
\end{theoremapp}

\textit{Proof:}  Consider the intermediate (possibly disconnected) polyomino 
after \textbf{Step 1}, call it $P^i$. There are many possible initial 
polyominoes which gives the same $P^i$, i.e., all polyominoes with the same 
configuration outside $G$.  They all map to ``almost'' the same resulting 
polyomino (the uncertainty implied in ``almost'' comes from the 
non-uniqueness of reconnecting in \textbf{Step 4} which will be discussed 
later).  Therefore, the mapping from domain to codomain is $M'$-to-one where 
$M'$ is bounded by the number of ways that at most $b^2$ cells can be 
connected within $G$.  Let the bound be $M$. 

On the other hand, \textbf{Step 4} of the drilling process is not unique.  
There might be  more than one way to connect those disconnected components 
to the punctured polyomino. But since $G$ is finite, the number of ways of 
reconnection is bounded above.  Let that bound be $C'$. 

Together with the drilling mapping $D$ defined above, we can write,
\[
s_n^{(0)} \le C'M \sum_{j = n - b^2}^{n + (b^2 - 2)}s_j^{(1)}.
\]
By the increasing monotonicity of $s_n^{(h)}$ over $n$ (proven in 
lemma (\ref{lem:incmono})),
\[
s_n^{(0)} \le (2b^2 -1)C'M s_{n + (b^2 - 2)}^{(1)}.
\]
Set $C = (2b^2 -1)C'M$ and the result follows. $\quad \blacksquare$ \\ 

Now consider the drilling of $h$ holes.
Denote by $\lceil c \rceil$ the ceiling of $c$, the smallest
integer greater than or equal to $c$.  Now, we could choose $h$ drilling 
locations from at least $\lceil n/b^2 \rceil$ grid squares.  To see this, 
consider first a polyomino with $n$ cells, where $n \le b^2$.  
The situation where we have the least number of drilling sites is when $n$ 
cells fall exactly within 1 grid square.  In this way, we only have 1 
possible grid square where we could drill holes.  Similarly, if we have 
$b^2n + w$ cells, where $n \ge 1$ and $0 \le w < b^2$, the minimal number 
of grid squares is when $b^2n$ cells fall into exactly $n$ grid squares and 
the other $w$ cells falls into another single grid. Then we have $n+1$ 
possible grid squares to drill holes.  Therefore we have at least 
$\lceil n/b^2 \rceil$ grid squares where we could drill holes.

Letting $H$ be a set of grid squares, let $\Phi_n^{[H]}$ denote the set of 
all polyominoes with $n$ cells and 
$|H|$ holes where there is a hole in each grid square of $H$.  Let 
$s_n^{[H]}$ denote its cardinality. One property of this set is that, 
for any $n_1, n_2 \in \textbf{$Z^+$}$ such that $n_1 \le n_2$,
\[
s_{n_1}^{[H]} \le s_{n_2}^{[H]}.
\]

\begin{theoremapp}
There exists a real constant $K$ such that for all $b \ge 5$,
\begin{equation}
\binom{\lceil n/b^2 \rceil}{h} s_n^{(0)} \le K s_{n+(b^2 - 2)h}^{(h)}, 
\qquad \forall b \le \lceil n/b^2 \rceil. 
\end{equation} 
\end{theoremapp}
\textit{Proof:} Place the polyomino in the grid system.  Let $A'$ be the set 
of all grid squares where we could drill holes.  From previous results, we 
know there are at least $\lceil n/b^2 \rceil$ elements in $A'$.  Truncate 
the set $A'$ with only the first $\lceil n/b^2 \rceil$ elements and call 
this set of drilling sites $A$.

Pick a subset $H \subseteq A$ such that 
$|H| = h$ $(\le \lceil n / b^2 \rceil)$ and drill holes one by one in 
each grid square in $H$, leading to a series of mappings:
\[
M :  \Phi_n^{(0)} \longmapsto \Phi_{m_1}^{(1)} \longmapsto \Phi_{m_2}^{(2)}
\longmapsto ... \longmapsto \Phi_{m_h}^{[H]} 
\]
where $m_i$'s are appropriate constants depending on each instance of 
operation and $ m_i \le n + (b^2 -2)h$, $ \forall i$.
So, from this mapping $M$, we have
\begin{equation} \label{eq:2.2.0}
s_n^{(0)} \le C^h s_{m_h}^{[H]} \le C^h s_{n+(b^2 -2)h}^{[H]}.  
\end{equation}
Hence 
\begin{eqnarray*}
\sum_{H \in A} s_n^{(0)} &\le& \sum_{H \in A} C^h s_{n+(b^2 -2)h}^{[H]} \\
\binom{\lceil n/b^2 \rceil}{h} s_n^{(0)} 
&\le& C^h  \sum_{H \in A} s_{n+(b^2 -2)h}^{[H]} \\
\binom{\lceil n/b^2 \rceil}{h} s_n^{(0)} 
&\le& K s_{n+(b^2 -2)h}^{(h)}.
\end{eqnarray*}
The last line arises since 
$\bigcup_{H \in A} \Phi_n^{[H]} \subseteq \Phi_n^{(|H|)} \equiv \Phi_n^{(h)}$ 
and put $K=C^h$.
$\quad \blacksquare$

\subsubsection{Surgery \label{sec:surgery}}

The surgery operation removes a linear sequence of cells inside the polyomino
and concatenates the sequence to the external boundary of the polyomino.  
Our objective is to join two holes thereby reducing the number of holes 
by one. 

We divide our domain into three classes:
 
(1) the set of polyominoes which have at least one hole with size one cell; 
denote this set by $\dot{\Phi}_n^{(h)}$ with cardinality $\dot{s}_n^{(h)},$ 
see figure~\ref{fig:surgery}(a);

(2) the set of polyominoes which are not in (1), and have three holes 
touching each  other at corners around a single cell; denote this set by 
$\ddot{\Phi}_n^{(h)}$ with cardinality $\ddot{s}_n^{(h)}$, 
see figure~\ref{fig:surgery}(b);
 
(3) the set of polyominoes which are not in (1) nor (2) (the general case), 
denote this set by $\tilde{\Phi}_n^{(h)} = 
\Phi_n^{(h)} \backslash (\dot{\Phi}_n^{(h)} \bigcup
\ddot{\Phi}_n^{(h)})$ with cardinality $\tilde{s}_n^{(h)}$.

\paragraph{The general case:}

First, let's look at polyominoes in $\tilde{\Phi}_n^{(h)}$.  Consider a 
polyomino $\alpha_n^{(h)} \in \tilde{\Phi}_n^{(h)}$.  Let $L$ be the set of 
all loop-free sequences\footnote{A loop-free sequence is a sequence of 
cells where each successive pair of cells are joined and no cell appears 
more than once in the whole sequence} of cells in $\alpha_n^{(h)}$, one end 
of which must touch the boundary of one hole and the other end must touch 
the boundary of another hole (or the exterior boundary). Define the length 
of the sequence to be the number of cells in the sequence. Let the set 
$Z$ be the set of sequences in $L$ that has minimum length.

Now pick one sequence $z \in Z$ and cut it out. This step will join two 
holes together (or join one hole and the exterior together).  Next, using 
the definition in section~\ref{sec:concat}, concatenate 
$\alpha_n^{(h)} \setminus cz \oplus D \oplus z$ ($D$ is a $3 \times 3$
polyomino block), and the resulting polyomino has one less hole and nine 
more cells than the original one.  So we have a mapping
\[
S_3 : \tilde{\Phi}_n^{(h)} \longmapsto \Phi_{n+9}^{(h-1)}
\]
This mapping is at most $n$-to-one because for any polyomino in the 
codomain, we could find at most $n$ locations to connect $z$ back.  As 
a result, we find that,
\begin{equation}
\tilde{s}_n^{(h)} \le n s_{n+9}^{(h-1)}.  \label{eq:sur1}
\end{equation}

\paragraph{Special cases:}

There are some subclasses of case (1) where we cannot apply the general 
surgery operation. One example is shown in figure~\ref{fig:nosurgery} 
\cite{Ren99}. In this example, the minimum  length of the loop-free sequence 
connecting two boundaries is two, but when we try  to remove any such 
minimum sequence, we will disconnect the polyomino.

To deal with case (1) polyominoes, we simply fill such a hole with a single 
cell. This satisfies our objective of reducing the number of holes by one.  
So we find a mapping
\[
S_1 : \dot{\Phi}_n^{(h)} \longmapsto \Phi_{n+1}^{(h-1)}.
\]
The mapping $S_1$ is $n'$ to 1 where $n'$ is less than $n$. To see this, 
consider a polyomino in the codomain.  To map back to the domain, we can 
choose any \textit{interior} cell to remove and the number of choices is 
obviously less than the total number of cells in the polyomino, so 
$n' \le n$.  Therefore,
\begin{equation}
\dot{s}_n^{(h)} \le n s_{n}^{(h-1)}.  \label{eq:sur2}
\end{equation}

Finally, consider polyominoes in $\ddot{\Phi}_n^{(h)}$.  The problem with 
these polyominoes is that when we try to remove the cell between the 3 
holes, we will inevitably join 3 holes together instead of joining 2.  
One way to get around this is to deliberately choose another sequence 
(which is also a cell in this class) in the set $Z$.  In particular, we 
choose to remove the cell $z' \in Z$ such that $z'$ is the top cell (in 
lexicographic ordering) of $Z$ which is not the problem cell.  Similarly 
to the general case, we find a mapping 
\[
S_2 : \ddot{\Phi}_n^{(h)} \longmapsto \Phi_{n+9}^{(h-1)}
\]
and hence
\begin{equation}
\ddot{s}_n^{(h)} \le n s_{n+9}^{(h-1)}.  \label{eq:sur3}
\end{equation}

Adding (\ref{eq:sur1}), (\ref{eq:sur2}), (\ref{eq:sur3}), we have
\[
s_n^{(h)} = \dot{s}_n^{(h)} + \ddot{s}_n^{(h)} + 
\tilde{s}_n^{(h)} \le n s_{n+9}^{(h-1)} 
+ n s_{n+9}^{(h-1)} + n s_{n}^{(h-1)}
\]
and hence:

\begin{theoremapp}
The number of polyominoes of area $n$ with $h$ holes is bounded above by $3n$ 
times the number of polyominoes of area (n+9) with one less hole.  That is,
\begin{equation}
s_n^{(h)} \le 3ns_{n+9}^{(h-1)}
\end{equation}
\end{theoremapp}

\subsection{Growth constants (by area)}

The following theorem proves the existence and equality of all the growth 
constants for $k$-punctured polyominoes for all finite $k$.

\begin{theoremapp}
There exists a constant $\beta_0$ such that for all $h \ge 0$ 
\begin{equation}
\lim_{n \to \infty} n^{-1} \log(s_n^{(h)}) = \log(\beta_0).
\end{equation}
\end{theoremapp}
\textit{Proof:} We use induction.  First, $\beta_0$ exists \cite{Ham61}.
Assume $\beta_h = \lim_{n \to \infty} n^{-1} \log(s_n^{(h)})$ exists. 
From the results of the concatenation and surgery operations, we have
\[
s_{n-m}^{(h)} s_m^{(1)} \le s_n^{(h+1)} \le 3ns_{n+9}^{(h)}.
\]
Choose some value $m$ such that $0 < s_m^{(1)} < \infty$, for example $m=8$.  
Take the logarithm, divide by $n$ and take the limit $n \to \infty$. 
This gives
\[
\log(\beta_h) \le  \log(\beta_{h+1}) \le \log(\beta_h)
\]
and hence $\beta_h = \beta_{h+1}$. Iterating from $\beta_0$ gives 
$\beta_0 = \beta_h$ for all finite $h$. $\quad \blacksquare$\\

\subsection{Relationships between critical exponents}

The following theorem establishes the relationships between critical 
exponents  should they exist.

\begin{theoremapp}
Assume for all $h$ $s_n^{(h)} \sim C_h n^{-\phi_h} \beta_0^n$ where $C_h$ 
is a $h$-dependent constant.  Then
\begin{equation}
\phi_h = \phi_0 - h.
\end{equation}
\end{theoremapp}
\textit{Proof:} From the results of the drilling operation, we have
\begin{equation}
C^{-h}\binom{\lceil n/b^2 \rceil}{h} s_{n-(b^2-1)}^{(0)} \le s_n^{(h)}.  
\label{eq:drillres} 
\end{equation}
Since
\begin{eqnarray*}
\binom{\lceil n/b^2 \rceil}{h} &\sim& \frac{1}{h!} \lceil n/b^2 \rceil^h,
\end{eqnarray*}
substituting the assumed asymptotic form $s_n^{(h)} 
\sim C_h n^{-\phi_h} \beta^n_h$ into (\ref{eq:drillres}), dividing by 
$\beta_0^{n}$, taking logarithms, dividing by $\log(n)$,
letting $n \to \infty$ and using the above result, we get
\begin{equation}
h - \phi_0 \le -\phi_h. 
\end{equation}
Next, from the result of the surgery operation, we have
\begin{eqnarray*}
s_n^{(h)} \le 3ns_{n + 9}^{(h-1)} & 
\le & (3n)3(n+9)s_{n+18}^{(h-2)} \le ... \\ 
& \le & (3n)3(n+9)...3(n+9h)s_{n+9(h-1)}^{(0)} 
\le [3(n + 9h)]^hs_{n+9h}^{(0)}. 
\end{eqnarray*}
Again by substituting the assumed asymptotic form, dividing by 
$\beta_0^{(n)}$, taking logarithms, dividing by $\log(n)$ and letting 
$n \to \infty$, we get
\begin{equation}
-\phi_h \le h - \phi_0.
\end{equation}
Hence $\phi_h = \phi_0 -h.$ $\quad \blacksquare$

\clearpage

\begin{table}
\caption{\label{tab:update}
The various `input' states and the `output' states (with corresponding 
weights) which arise as the boundary line is moved in order to include 
one more vertex of the lattice.}
\begin{center}
\begin{tabular}{ccc}  \hline  \hline
Input & \multicolumn{2}{c}{Outputs} \\ \hline
`00' & `00' & $x^2$`12'  \\
`01'/`10'  & $x$`01' & $x$`10' \\
`02'/`20'  & $x$`02' & $x$`20' \\
`11'/`22' & `$\overline{00}$' &  \\
`21' & `00'  \\
`12' & `accumulate'  \\ \hline \hline
\end{tabular}
\end{center}
\end{table}

\begin{table}
\caption{\label{tab:fit} A fit to the asymptotic form 
$p_{2n}^{(0)} \sim \mu^2 n^{-\frac{5}{2}}[c_0 + c_1/n + c_2/n^2 
+ c_3/n^3 + \cdots]$ for the number of SAP enumerated by perimeter$.$
Estimates of the amplitudes $c_0, c_1, c_2, c_3.$}\begin{center}
\begin{tabular}{|c|cccc|}
\hline \hline
$n$ & $c_0$ & $c_1$ & $c_2$ & $c_3$  \\ \hline
 20 &  0.09940085 & -0.02745705 &  0.02476376 &  0.11822181 \\
 21 & 0.09940118  &-0.02747548  & 0.02511347 &  0.11601107 \\
 27 &  0.09940177 & -0.02751355 &  0.02593211 &  0.11011880 \\
 28 &  0.09940179 & -0.02751510 &  0.02597236 &  0.10977030\\
 29 &  0.09940180 & -0.02751619 &  0.02600168 &  0.10950667\\
 30 &  0.09940181 & -0.02751694 &  0.02602273 &  0.10931043\\
 31 &  0.09940182 & -0.02751745 &  0.02603734 &  0.10916929\\
 32 &  0.09940182 & -0.02751777 &  0.02604692  & 0.10907354\\ 
 33 &  0.09940182 & -0.02751795 &  0.02605254 &  0.10901552\\
 34 &  0.09940182 & -0.02751802 &  0.02605500 &  0.10898929\\
 35 &  0.09940182 & -0.02751802 &  0.02605494 &  0.10898993\\
 36 &  0.09940182 & -0.02751796 &  0.02605285 &  0.10901358\\
 37 &  0.09940182 & -0.02751785 &  0.02604913 &  0.10905699\\
 38 &  0.09940182 & -0.02751771 &  0.02604408 &  0.10911757\\
 39 &  0.09940182 & -0.02751755 &  0.02603796 &  0.10919302\\
 40 &  0.09940182 & -0.02751736 &  0.02603097 &  0.10928158\\
 41 &  0.09940182 & -0.02751717 &  0.02602327 &  0.10938160\\
 42 &  0.09940181 & -0.02751696 &  0.02601500 &  0.10949174 \\
 43 &  0.09940181 & -0.02751675 &  0.02600629 &  0.10961079 \\
 44 &  0.09940181 & -0.02751653 &  0.02599720 &  0.10973796 \\
 45 &  0.09940181 & -0.02751631 &  0.02598785 &  0.10987195 \\
\hline \hline
\end{tabular}
\end{center}
\end{table}

\clearpage
\begin{table}
\caption{\label{tab:fit2} A fit to the asymptotic form $a_n^{(1)} \sim
\eta^n n[c_0 + c_1/n^{\frac{1}{2}} + c_2/n + c_3/n^{\frac{3}{2}}
 + c_4/n^2 + \cdots]$ for the number of 1-punctured staircase polygons 
enumerated by area$.$
Estimates of the amplitudes $c_0, c_1, c_2, c_3, c_4.$}\begin{center}
\begin{tabular}{|c|ccccc|}
\hline \hline
$n$ & $c_0$ & $c_1$ & $c_2$ & $c_3$& $c_4$  \\ \hline

 94 & 6.83279$\times 10^{-3}$& -1.86263$\times 10^{-2}$
    & -2.63249$\times 10^{-2}$&  2.95964$\times 10^{-2}$
    &  8.60593$\times 10^{-2}$\\
 95 & 6.83258$\times 10^{-3}$& -1.86182$\times 10^{-2}$
     & -2.64419$\times 10^{-2}$&  3.03465$\times 10^{-2}$
     &  8.42557$\times 10^{-2}$\\
 96 & 6.83256$\times 10^{-3}$& -1.86173$\times 10^{-2}$
     & -2.64559$\times 10^{-2}$&  3.04363$\times 10^{-2}$
     &  8.40388$\times 10^{-2}$\\
 97&  6.83239$\times 10^{-3}$& -1.86109$\times 10^{-2}$
     & -2.65488$\times 10^{-2}$&  3.10389$\times 10^{-2}$
     &  8.25743$\times 10^{-2}$\\
 98 & 6.83235$\times 10^{-3}$& -1.86092$\times 10^{-2}$
     & -2.65738$\times 10^{-2}$&  3.12014$\times 10^{-2}$
     &  8.21773$\times 10^{-2}$\\
 99&  6.83222$\times 10^{-3}$& -1.86040$\times 10^{-2}$
     & -2.66501$\times 10^{-2}$&  3.17010$\times 10^{-2}$
     &  8.09504$\times 10^{-2}$\\
100 & 6.83217$\times 10^{-3}$& -1.86019$\times 10^{-2}$
     & -2.66815$\times 10^{-2}$&  3.19076$\times 10^{-2}$
     &  8.04405$\times 10^{-2}$\\
101 & 6.83206$\times 10^{-3}$& -1.85976$\times 10^{-2}$
     & -2.67457$\times 10^{-2}$&  3.23325$\times 10^{-2}$
     &  7.93864$\times 10^{-2}$\\
102 & 6.83200$\times 10^{-3}$& -1.85953$\times 10^{-2}$
     & -2.67801$\times 10^{-2}$&  3.25612$\times 10^{-2}$
     &  7.88161$\times 10^{-2}$\\
103 & 6.83191$\times 10^{-3}$& -1.85916$\times 10^{-2}$
     & -2.68355$\times 10^{-2}$&  3.29313$\times 10^{-2}$
     &  7.78885$\times 10^{-2}$\\
104 & 6.83185$\times 10^{-3}$& -1.85893$\times 10^{-2}$
     & -2.68708$\times 10^{-2}$&  3.31690$\times 10^{-2}$
     &  7.72899$\times 10^{-2}$\\
105 & 6.83177$\times 10^{-3}$& -1.85860$\times 10^{-2}$
     & -2.69197$\times 10^{-2}$&  3.34985$\times 10^{-2}$
     &  7.64560$\times 10^{-2}$\\
106 & 6.83171$\times 10^{-3}$& -1.85837$\times 10^{-2}$
     & -2.69553$\times 10^{-2}$&  3.37400$\times 10^{-2}$
     &  7.58417$\times 10^{-2}$\\
107 & 6.83164$\times 10^{-3}$& -1.85809$\times 10^{-2}$
     & -2.69988$\times 10^{-2}$&  3.40364$\times 10^{-2}$
     &  7.50842$\times 10^{-2}$\\
108 & 6.83159$\times 10^{-3}$& -1.85786$\times 10^{-2}$
     & -2.70333$\times 10^{-2}$&  3.42727$\times 10^{-2}$
     &  7.44775$\times 10^{-2}$\\
109 & 6.83153$\times 10^{-3}$& -1.85761$\times 10^{-2}$
     & -2.70731$\times 10^{-2}$&  3.45466$\times 10^{-2}$
     &  7.37709$\times 10^{-2}$\\
110 & 6.83148$\times 10^{-3}$& -1.85739$\times 10^{-2}$
     & -2.71063$\times 10^{-2}$&  3.47758$\times 10^{-2}$
     &  7.31769$\times 10^{-2}$\\
111 & 6.83142$\times 10^{-3}$& -1.85716$\times 10^{-2}$
     & -2.71427$\times 10^{-2}$&  3.50291$\times 10^{-2}$
     &  7.25171$\times 10^{-2}$\\
112 & 6.83137$\times 10^{-3}$& -1.85696$\times 10^{-2}$
     & -2.71747$\times 10^{-2}$&  3.52522$\times 10^{-2}$
     &  7.19335$\times 10^{-2}$\\
113 & 6.83132$\times 10^{-3}$& -1.85675$\times 10^{-2}$
     & -2.72077$\times 10^{-2}$&  3.54832$\times 10^{-2}$
     &  7.13266$\times 10^{-2}$\\
\hline \hline
\end{tabular}
\end{center}
\end{table}
\clearpage
\begin{table}
\caption{\label{tab:amps} Amplitude estimates appearing in the
asymptotic form of the
coefficients of the various models considered. The order of the
term associated with the amplitude $c_i$ varies from model to model, and
is given in the text for each model. The various connective constants are:
$\eta=2.30913859330, \kappa = 3.97094397, \mu = 2.63815853034,
\tau = 4.062591 $. All numbers quoted are expected to have errors
only in the last quoted digit. $k$ is the number of punctures.}
The $n^{th}$ coefficient is given by Prefactor$\times \sum_{i\ge0}
c_i/n^{f(i)}.$ See the text for the problem dependent value of $f(i).$
\begin{center}
\begin{tabular}{|c|c|c|cccc|}
\hline \hline
Model, & $k$ & Prefactor & $c_0$ & $c_1$ & $c_2$ & $c_3$  \\
Parameter. &  &  &  &  &  &   \\ \hline
Staircase & 0 &$\eta^n$ & 0.12881579 & $O(1.5996^{-n})$ &  &   \\
polygons & 1  & $\eta^n n$  & 0.006831  &-0.0185   &-0.028 & 0.04  \\
by area  & 2   &$ \eta^n n^2$ & $7.87 \times 10^{-5}$  
         & -0.00043 & 0.0010  & -0.007  \\
& 3  & $\eta^n n^3$  & $6.04 \times 10^{-7}$  & $-5.03 \times 10^{-6}$ 
     & 0.000031 & -0.00022 \\ \hline
Staircase & 0 &$4^n n^{-\frac{3}{2}}/\sqrt{\pi}$ & $1/4$ 
          &$3/32$ &$25/512$  & $64/4096$   \\
polygons & 1  & $4^n$  & 0.0147  &$ -0.19$ & 1.4  &\\
by       & 2  & $4^n n^{\frac{3}{2}}$& $8.0\times 10^{-4}$ 
         & $-2\times 10^{-2}$    & 0.2  &  \\
perimeter& 3  & $4^n n^3 $  & $3.0\times 10^{-6}$ 
         &  $-1\times 10^{-3}$  & 0.02 &\\ \hline
SAP  & 0 & $\kappa^n/n$  & 0.408105  &$-0.5467$   &0.626 &$ -3$ \\
by   & 1 &  $\kappa^n$  & 0.000975  &$-0.0097 $  &$-0.04$  &\\
area & 2 &  $\kappa^n n$  &0.00000118   &$-0.000019$   &$-0.0001$ &\\
  & 3  & $\kappa^n n^2$  &$ 1.0\times 10^{-9} $  
       &$-2\times 10^{-8}$ &$-3 \times 10^{-7} $  & \\ \hline
SAP & 0 & $\mu^{2n}n^{-\frac{5}{2}}$  & 0.0994018&$ -0.02751$ &0.0255 &0.12 \\
by       & 1  & $\mu^{2n}/n$   & 0.001444 &$ -0.00843$  &0.0078   & 0.026  \\
perimeter& 2 &$\mu^{2n}n^{\frac{1}{2}}$& 0.000011 &$ -0.0001$  & 0.0005   & \\
 & 3  & $\mu^{2n} n^2$  &$1\times 10^{-7} $  &   &  & \\ \hline

Polyominoes&--  & $\tau^n/n$  & 0.31660  & $-0.233$  &0.62   &$-2.5$ \\ \hline

Punctured  &1 & $\kappa^n$  &0.00922   &$ -0.107$  & 0.30 &  \\
polyominoes&2  & $\kappa^n n$  &0.000104   &$ -0.0022 $ &0.009   &   \\
           &3  & $\kappa^n n^2$  &0.0000008   &$ -0.00002$  &0.0002 &   \\
\hline \hline
\end{tabular}
\end{center}
\end{table}

\clearpage

\setcounter{figure}{0}
\def\thefigure{A\arabic{figure}}

\begin{figure}
 \begin{center}
   \includegraphics[height=8cm]{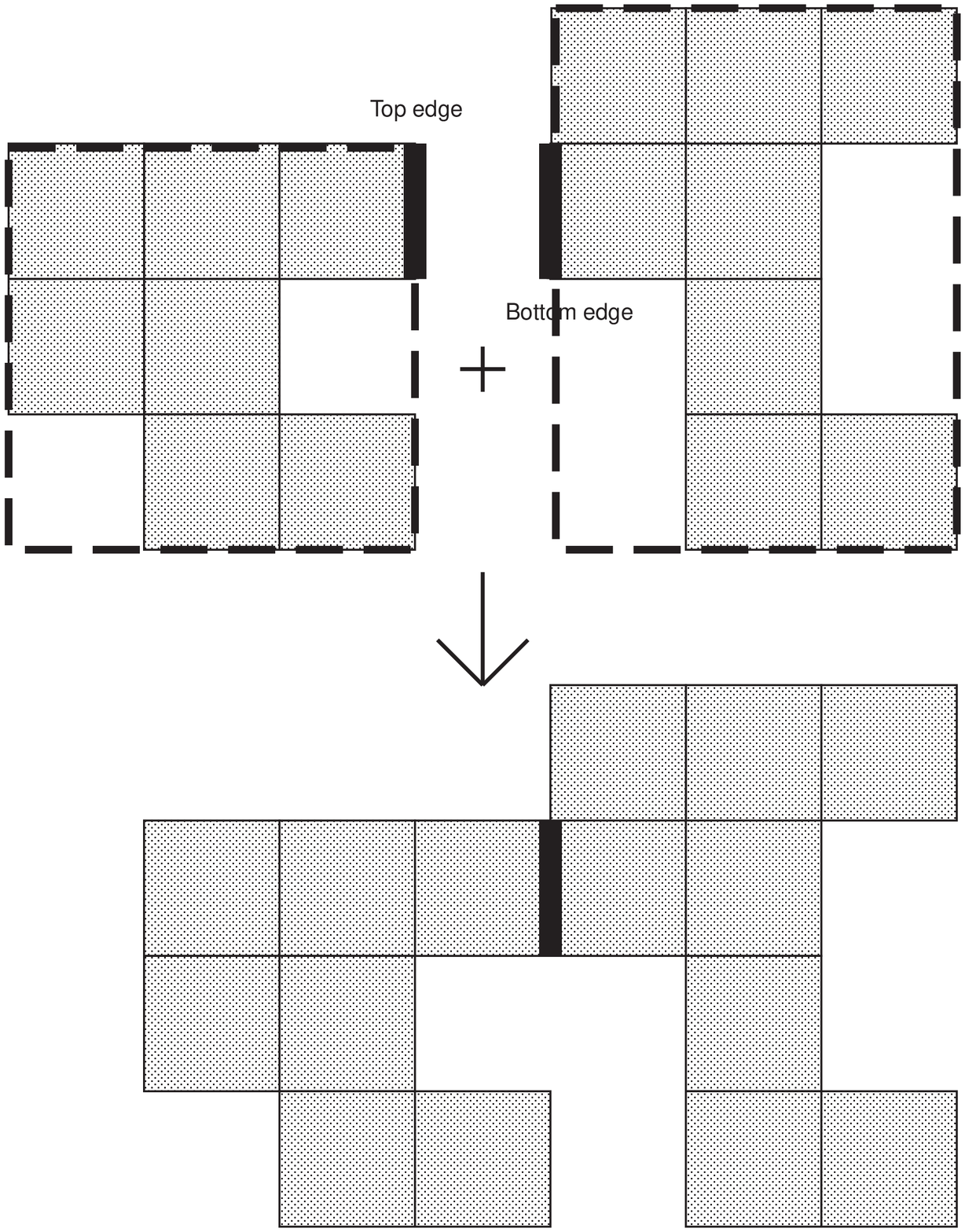}
   \caption{\label{fig:concat} 
   Concatenation operation}
 \end{center}
\end{figure}

\begin{figure}
 \begin{center}
   \includegraphics{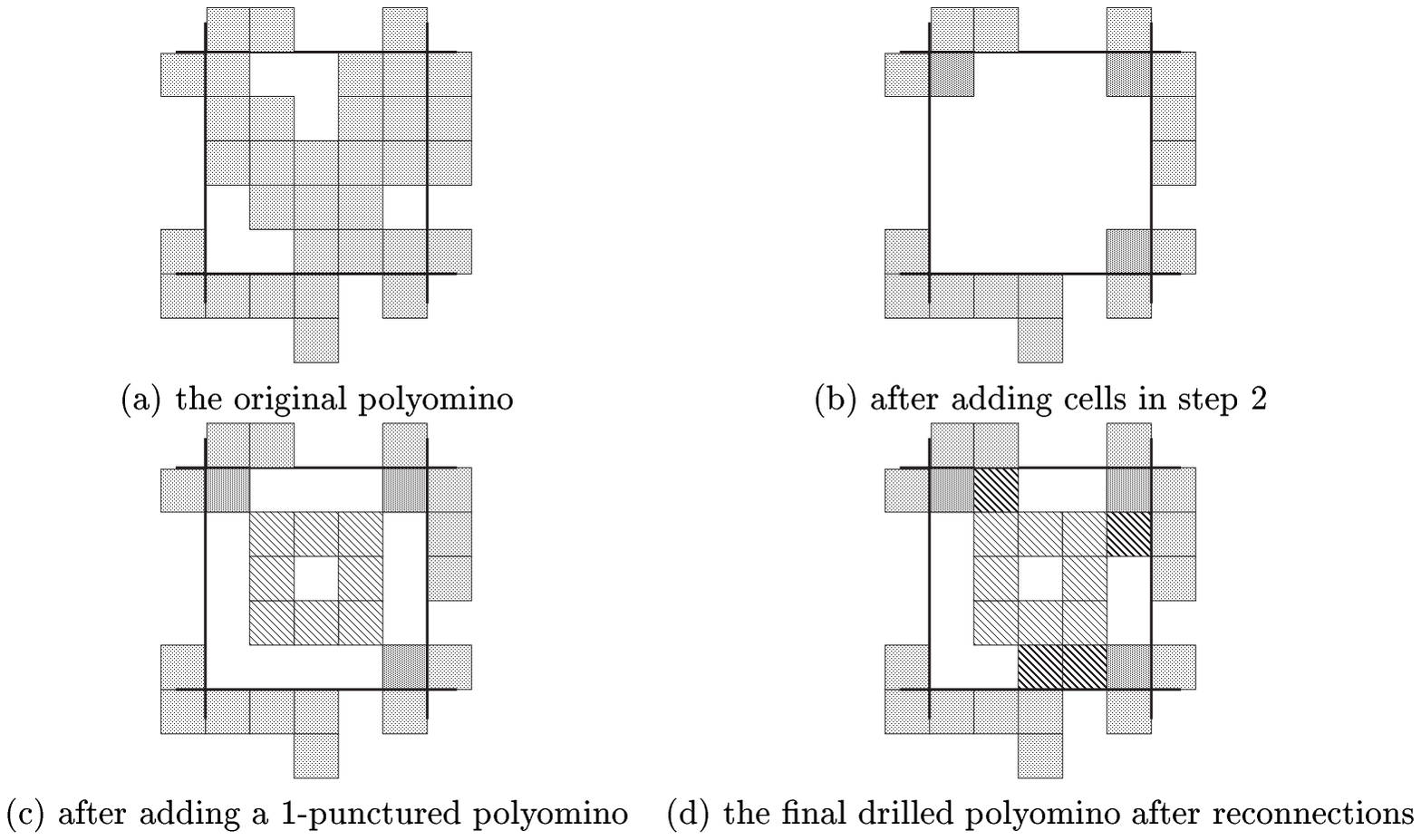}
 \end{center}
\caption{\label{fig:drill}
Drilling operation: (a) The original polyomino.  After step 1, all the
cells within the grid square are removed.  In step 2, we reconnect all the 
disconnected components around the corner (shown in (b)).  Then, in step 3, 
a 1-punctured polyomino is placed at the centre of the grid square (shown 
in (c)). Finally, all the disconnected pieces outside the grid square are 
reconnected to the punctured polyomino (shown in (d)).}
\end{figure}

\begin{figure}
 \begin{center}
   \includegraphics{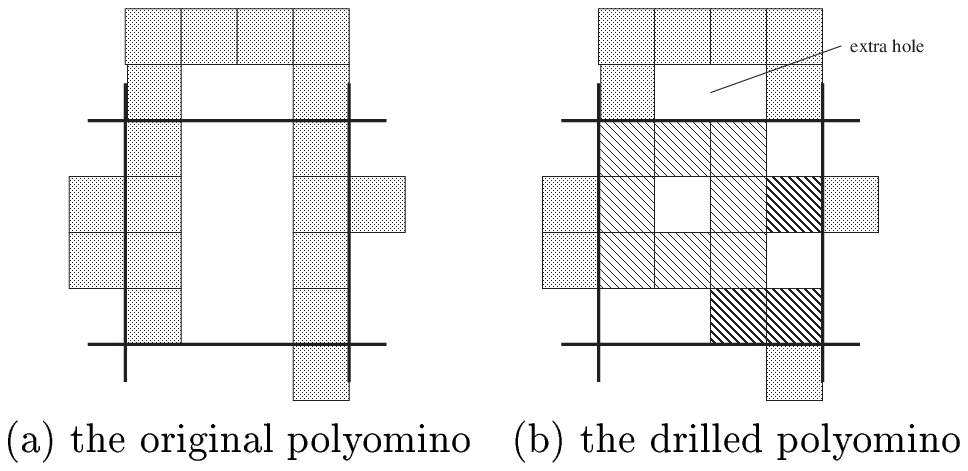}
 \end{center}
\caption{\label{fig:exdrill}
An illustration of a grid square that is too small.}
\end{figure}

\begin{figure}
\begin{center}
   \includegraphics{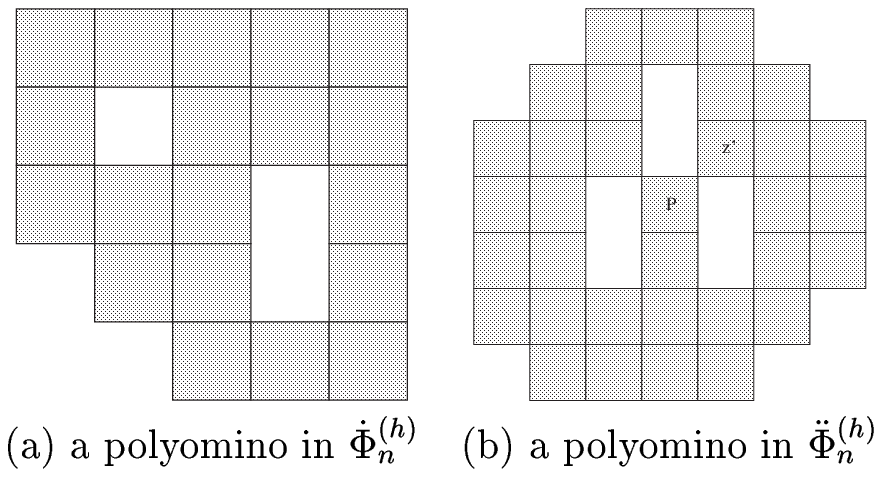}
\end{center}
\caption{\label{fig:surgery}
An illustration of two special cases for the surgery operation.  
In (b), $P$ denotes the problem cell whose removal will join 3 holes together.
In this case we choose to remove $z'$}
\end{figure}

\begin{figure}
 \begin{center}
   \includegraphics[height=4cm]{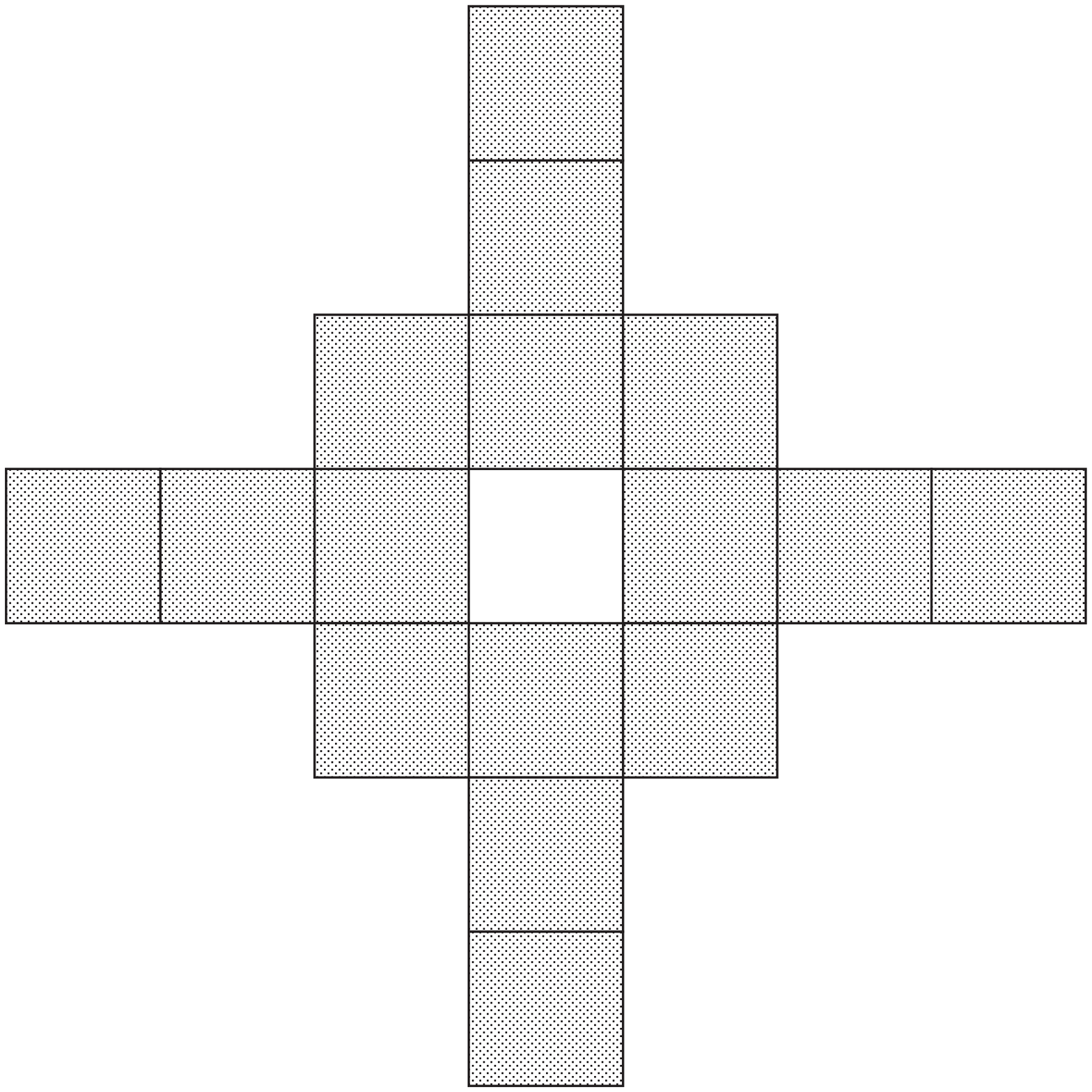}
 \end{center}
\caption{\label{fig:nosurgery}
A polyomino in $\dot{\Phi}_n^{(h)}$ where we cannot apply the
general surgery operation}
\end{figure}

\end{document}